\documentclass{article}
\usepackage[utf8]{inputenc}
\usepackage{geometry}
\usepackage{amsmath}
\usepackage{amsfonts}
\usepackage{mathrsfs}
\usepackage{mathtools}
\usepackage{xcolor}
\usepackage{graphicx}
\usepackage{comment}
\usepackage{fixmath}
\usepackage{bm}
\usepackage{hyperref}
\usepackage{arydshln}
\usepackage{booktabs}

\usepackage[round]{natbib}
\title{Stochastic Actor Oriented Model with Random Effects}
\author{Giacomo Ceoldo, Tom A.B. Snijders, Ernst C. Wit}
\date{}

\begin{document}

\maketitle

\begin{abstract}
The stochastic actor oriented model (SAOM) is a method for modelling social interactions and social behaviour over time. It can be used to model drivers of dynamic interactions using both exogenous covariates and endogenous network configurations, but also the co-evolution of behaviour and social interactions. In its standard implementations, it assumes that all individual have the same interaction evaluation function. This lack of heterogeneity is one of its limitations. The aim of this paper is to extend the inference framework for the SAOM to include random effects, so that the heterogeneity of individuals can be modeled more accurately.

We decompose the linear evaluation function that models the probability of forming or removing a tie from the network, in a homogeneous fixed part and a random, individual-specific part. We extend the algorithm so that the variance of the random parameters can be estimated with method of moments. Our method is applicable for the general random effect formulations. We illustrate the method with a random out-degree model and show the parameter estimation of the random components, significance tests and model evaluation. We apply the method to the Kapferer's Tailor shop study. It is shown that a random out-degree constitutes a serious alternative to including transitivity and higher-order dependency effects.
\end{abstract}

\section{Introduction}
\label{section:Introduction}

Modeling the behavior of a group of people when interactions between them emerge and dissolve, is one the most ambitious goals of social network analysis.
The first overview of statistical models used to analyze social network data appeared in the 1990s \citep{wasserman1994social}, following pioneering work on the quantitative study of social networks in the 1970s \citep[e.g.][]{holland1977dynamic,holland1977method}.
During this development, various basic continuous-time Markov chains were proposed to describe the evolution of a social network under a number of simplifying assumptions \citep{wasserman1980analyzing}.

This formed the basis for the stochastic actor oriented model for network change \citep{snijders1996stochastic,snijders2001statistical,snijders2017stochastic}, which has become known as the \emph{SAOM}. An important alternative to the SAOM is the \emph{Temporal Exponential Random Graph Model} (\emph{TERGM}) described in \cite{hanneke2010discrete} and \cite{krivitsky2014separable}.
An extensive comparison of the two models is in \cite{block2018change} and \cite{block2019forms}. \emph{Relational Event Models} (\emph{REM}; \citealp{butts20084}; \citealp{perry2013point}) and \emph{Dynamic Network Actor Models} (\emph{DyNAM}; \citealp{stadtfeld2011analyzing}; \citealp{stadtfeld2017dynamic}) can also be used for temporal network data. However, these approaches are used to model instantaneous interaction events, such as emails, citations, or emergency calls, whereas SAOM and TERGMs are models for networks of ties that endure over time.

The SAOM parametrizes the evolution of the stochastic network of relations.
The model is used in longitudinal (or panel) network studies, that are observational studies where, in its simplest form, a relational network on a fixed set of actors is observed at two or more time points.
We assume here that the network is directed.
An example is a friendship network between students in a classroom.
The evolving set of relations is viewed as a stochastic process in continuous time on the outcome space of directed graphs, assuming that changes, when they occur, consist of a change of only one tie variable.
The process is observed at discrete time points, and unobserved in between.
The SAOM was first introduced in \cite{snijders2001statistical}.
For a review of the model, including a discussion of various generalizations, see \cite{snijders2017stochastic}.
An extensive discussion of the model is in the manual of the package \emph{RSiena} \citep{manualRsiena}.

The SAOM combines a parametrization for the temporal component that specifies when and who has the possibility to change a tie in the network, with a model for the probabilities of changes in the outcome space, which are specified by a different set of parameters.
Our generalization concerns the latter component, whose parameters quantify how much a given configuration of the network, called effect, influences the choice of forming or dissolving ties.
These parameters are shared by all actors, so differences between individual rules
of network change have to be defined by means of observed covariates.

The study of social behaviour often either focuses on modelling the choices of heterogeneous actors in a relatively simple environment, or on modelling the choices of relatively homogeneous actors in a complex environment.
The generalization we are proposing is to allow some parameters to be random, to model more accurately the heterogeneity of the actors in the choices they make. The heterogeneity of the actors is then parametrized by the variance of the random parameters.

The first idea of using random effects in the stochastic actor oriented model is in \citet[Chapter 4]{schweinberger2007random}, where the estimation method, frequentist or Bayesian, is based on the likelihood function.
However, this method is not implemented in \emph{RSiena}.
The algorithm that will be explained here is completely different as it is based on the method of moments, which is the estimation algorithm most used in the SAOM.
In \cite{schweinberger2020statistical}, a related model is developed in which some parameters are shared by members of the same group while varying between groups, but its Bayesian estimation method is also based on the likelihood function.
A similar approach for relational event model is in \cite{dubois2013hierarchical}, where multiple sequences of relational events are observed, and the random effects are defined across sequences.
In exponential random graph models, random effects defined across actors have been introduced in 
\cite{thiemichen2016bayesian}, and for models with independent ties in \cite{van2004p2}, but we are not aware of the inclusion of random effects in temporal ERGMs.
Dynamic network actor models with random effects defined across actors have been developed in \cite{uzaheta2023random}.

In Section~\ref{section:SAOMRandomOutEffects}, the SAOM is introduced and the inclusion of random effects in the evaluation function is discussed.
As an example we describe extensively a model with random out-degree.
Various models for the variance of the random effects are also discussed. The estimation method described in Section~\ref{sec:estimation} is an extension of the Robbins-Monro algorithm that is used in the standard SAOM, the main difference being that the positive definiteness of the variance parameters must be enforced.
Section~\ref{section:ModelEvaluationComputationStandardErrors} explains how to test hypotheses on the variance parameters and how to compute the standard error of the estimates. The theory is described in detail for the hypothesis of overdispersion, i.e., heterogeneity in the out-degrees of the actors not explained by the model with fixed effects only. It is then shown how to compute the standard error of the estimated parameters in this model. Various generalizations of the model evaluation procedure are  briefly discussed, important examples are models with multiple random effects and reparametrizations in the variance parameter. In Section~\ref{section:EstimationKapferer} we apply the derived method to a tailor shop social network, reanalyzing a study by \cite{kapferer1972strategy}.
The study focuses on the changes in a social network of 39 individuals in a tailor shop in Zambia at the start and the end of a 7-month period.
Previous analyses have shown a marked effect of various transitive effects in the interactions between the individuals.
Various models, with and without random out-degree, are estimated and compared with the algorithm and model evaluation procedures derived in previous sections.
Several parameters are interpreted in detail to discuss how, and possibly why, the inclusion of the random out-degree affects the estimation of the other parameters.

In the rest of the paper, quantities written in boldface denote multivariate objects, like vectors (e.g., $\bm{\beta}$ or $\bm{s}$), matrices (written with uppercase letters, e.g., $\bm{\Sigma}$), networks ($\bm{x}$, or $\bm{X}$ when viewed as a random variable), sets (e.g., $\bm{\mathcal{X}}$), or functions (with vector codomain, e.g., $\bm{g}(\bm{x})$, or with matrix codomain, e.g., ${\bm{D}}(\bm{\beta})$).
With $\bm{\beta}^\top$ is the denoted the transpose of a vector (that is a row-vector), $\bm{\Gamma}^\top$ denotes the transpose of a matrix.

\section{Stochastic actor oriented model with random effects}
\label{section:SAOMRandomOutEffects}

In this section we describe the \emph{Stochastic Actor Oriented Model} (\emph{SAOM}) with the random effect generalization that we propose.
The estimation method will be discussed in detail in Section~\ref{sec:estimation}.

\subsection{Stochastic actor oriented model}

We first describe the evolution of the process in detail in a simplified set-up.
We assume that a simple directed graph with a node set consisting of $N$ actors is observed at $M=2$ time points $t_1$ and $t_2$.
The \emph{state space} of the process (possible values for the dependent variable) is $\bm{\mathcal{X}} \cong \{0,1\}^{N(N-1)}$, which can be represented as the set of binary matrices with values 0 in the diagonal.
States $\bm{x}\in\bm{\mathcal{X}}$ are adjacency matrices with elements $x_{ij}=1$ in row $i$ and column $j$ if the tie $(ij)$ is in the graph, otherwise $x_{ij}=0$; moreover, $x_{ii}=0$ for all $i$.
The observed networks are $\bm{x}(t_1)$ and $\bm{x}(t_2)$, respectively.
The unobserved evolution of the network between $t_1$ and $t_2$ is the right continuous random function $t\mapsto \bm{X}(t)$ such that $\bm{X}(t)\in \bm{\mathcal{X}}$, $\bm{X}(t_1)=\bm{x}(t_1)$ and $\bm{X}(t_2)=\bm{x}(t_2)$.
If the process changes at time $t$, an actor, denoted here by $i\in \{1,...,N\}$, has the opportunity to add or remove one outgoing tie, or to leave the configuration as it is.
Therefore, if $\bm{x}_0=\bm{x}(t^-)$ is the value of the network (state of the process) immediately before the possible change, the process can jump to the state $\bm{x}\in \bm{\mathcal{A}}_i(\bm{x}_0)\subset \bm{\mathcal{X}}$, where the \emph{adjacency set} $\bm{\mathcal{A}}_i(\bm{x}_0)$ is the set of networks that differ from $\bm{x}_0$ in at most one tie that starts from $i$.

The temporal component, which is not changed in our generalization of the stochastic actor oriented model, is now briefly discussed.
The generative process is assumed to be \emph{Markovian in continuous time}, with exponentially distributed time between transitions.
If a transition happens at time $t_0\in(t_1, t_2)$, the next transition can occur at a random time computed in the following way.
For all actors $j$ in $\{1,...,N\}$, random times $\Delta T_j$ are sampled independently from an exponential distribution with parameter $\lambda_j$.
Then the next transition will happen at time $t = t_0 + \min_j(\Delta T_j)$, if $t < t_2$ (otherwise the process end), and the \emph{focal actor} that has the possibility to flip one of the outgoing ties is $i = \text{argmin}_j(\Delta T_j)$.
Often $\lambda_j=\lambda$ for all $j$, but $\lambda_j$ can also depend on \emph{temporal effects} depending on covariates and/or network configurations.

The model for the opportunity of transitions in the state space is now introduced.
In the stochastic actor oriented model, $\bm{X}(t)$ is a random variable with probability distribution conditional on actor $i$ making a network change
\begin{equation}
P(\bm{X}(t)=\bm{x} \mid i,\bm{X}(t^-)=\bm{x}_0)
= \frac{\exp({f_{i}(\bm{x})})I(\bm{x}\in \bm{\mathcal{A}}_i(\bm{x}_0))}{\sum_{\bm{x}'\in\bm{\mathcal{A}}_i(\bm{x}_0)}\exp({f_{i}(\bm{\bm{x}'})})},
    \label{equation:transitionProbabilityUtility}
\end{equation}
where $I:\{\texttt{FALSE},\texttt{TRUE}\}\to\{0,1\}$ is the \emph{indicator function} that returns $1$ when the relation inside it is satisfied ($\texttt{TRUE}$), and $f_{i}:\bm{\mathcal{X}}\to\mathbb{R}$ is the \emph{linear evaluation function} of actor $i$
\begin{equation}
    f_{i}(\bm{x})
    =\bm{\beta}^\top \bm{s}_i(\bm{x})=\sum_{k=1}^p \beta_k s_{ik}(\bm{x}),
    \label{equation:linearUtilityFunction}
\end{equation}
parametrized by $\bm{\beta}\in\mathbb{R}^p$ that weights the $p$ dimensional vector of statistics $\bm{s}_{i}(\bm{x})$, which contains information on the ``position" of the focal actor $i$ in the network.
As $\bm{x}_0\in \bm{\mathcal{A}}_i(\bm{x}_0)$ for all $\bm{x}_0$, the ``trivial" transition in which the network does not change has always a positive probability.
The components $s_{ik}(\bm{x})$ are called \emph{effects}. Some examples are the following:
\begin{itemize}
    \item \emph{out-degree} (\emph{density}) \emph{effect}: number of ties starting from the focal actor $x_{i+}=\sum_jx_{ij}$, 
    \item \emph{reciprocity effect}: number of reciprocated ties $\sum_jx_{ij}x_{ji}$,
    \item \emph{transitive triplet effect}: number of ordered pairs $(j,h)$ of actors such that the three ties $(ij)$, $(ih)$ and $(hj)$ are all present in the graph, computed as $\sum_{j,h}x_{ij}x_{ih}x_{hj}$,
    \item \emph{out-degree related popularity effect}: sum of the out-degrees of the actors to whom $i$ is tied, computed as $\sum_jx_{ij}x_{j+}$.
\end{itemize}
Effects can also depend on actor covariates or dyadic covariates.

\subsection{Random effects in the SAOM}

The generalization proposed in this paper, is to decompose the evaluation function in a \emph{fixed} and a \emph{random part}, so that the actors in the network can have individual parameters.
The linear evaluation function (\ref{equation:linearUtilityFunction}) is replaced with
\begin{equation}
f_{i}(\bm{x}) =
\bm{\beta}^\top \bm{s}_i(\bm{x}) + \bm{b}_{i} ^\top\bm{r}_{i}(\bm{x}) =
\sum_{k=1}^p \beta_k s_{ik}(\bm{x}) + \sum_{h=1}^q b_{ih} r_{ih}(\bm{x}),
\label{equation:linearRandomUtilityFunction}
\end{equation}
where $\bm{b}_{i}\sim \mathcal{N}(\bm{0},\bm{\Sigma})$ is the vector of \emph{random parameters} for actor $i$, the between-actor \emph{variance parameter} $\bm{\Sigma}$ is a $q$ dimensional positive definite matrix, $q$ is the number of \emph{random effects} $r_{ih}(\bm{x})$, which are statistics like the ones given above.
The values $s_{ik}(\bm{x})$ are now called \emph{fixed effects}. 
It is possible that some statistics are both fixed and random effects.
In applications it is wise to avoid the use of random effects not present in the model also as fixed effects, as this can bias the estimate of $\bm{\Sigma}$.
The parameter $\bm{\Sigma}$ is used to model the variability for the individuals in considering the effects $\bm{r}_{i}(\bm{x})$ important in their evaluation function.

To make the discussion more concrete we will illustrate the approach using a simple example in which the out-degree is the only random effect.
The parameter for the out-degree effect balances creation of new ties with termination of existing ties and plays a role similar to the intercept in logistic regression.
The \emph{stochastic actor oriented model with random out-degree} is defined by (\ref{equation:linearRandomUtilityFunction}) with $q=1$ and $s_{i1}(\bm{x}) = r_{i1}(\bm{x}) = x_{i+}$, yielding the evaluation function
\begin{equation}
f_{i}(\bm{x})= (\beta_1 + b_i) s_{i1}(\bm{x}) +
\sum_{k=2}^p \beta_k s_{ik}(\bm{x}),
\label{equation:RandomOutDegUtilityFunction}
\end{equation}
where  $b_i\sim\mathcal{N}(0,\sigma^2)$.
In addition to the fixed parameter $\bm{\beta}$, the variance $\sigma^2$ has to be estimated, therefore the parameter space is $\mathbb{R}^p\times (0,\infty)$.

The variance parameter $\sigma^2$ for the random out-degree can also partially account for the bias that actors can have when nominating ties in a questionnaire (most common way to collect datasets analysed with the SAOM), as discussed in \citet{feld2002detecting}. This is called \emph{expansiveness bias}, as it refers to the fact that individuals can be different in considering a relation strong enough to be reported. The random out-degree can control this bias by incorporating its effect into $\sigma^2$. 


When more than one random effect is included in the model, different parameter spaces can be used for $\bm{\Sigma}$.
In the \emph{unrestricted} model for the variance $\bm{\Sigma}\in\mathbb{S}^+_q$, so the parameter space, which is the set of $q$ dimensional positive definite matrices, is $q(q+1)/2$ dimensional.
Other models can be specified by considering a smaller parameter space: restricting the number of parameters that are free to vary.
An important example is the \emph{unrestricted diagonal} variance $\bm{\Sigma}=\mathrm{diag}(\sigma_h^2)\in \text{diag}(\mathbb{S}_q^+)\cong (0,\infty)^q$, with a $q$ dimensional parameter space.

Other examples for restricted models are the following:
\begin{itemize}
    \item \emph{scalar}: $\bm{\Sigma}=\sigma^2\bm{I}$, $1$ degree of freedom;
    \item \emph{two-parameters}: $\bm{\Sigma}=\sigma^2\bm{I}+\varrho(\bm{J}-\bm{I})$, $\bm{J}$ is the matrix of ones, $ \varrho \in (-\sigma^2/(q-1),\sigma^2)$, $2$ degrees of freedom;
    \item \emph{autoregressive-1}: $\bm{\Sigma}=(\sigma_{hh'})$ such that $\sigma_{hh'} = \phi^{|h-h'|} \sigma^2/(1-\phi^2)$, $\phi\in (-1,1)$, 2 degrees of freedom;
    \item \emph{block-diagonal}: $\bm{\Sigma}=\text{blockdiag}(\bm{\Sigma}_h)$, where each $\bm{\Sigma}_h$ can have its own model, for example one of the models described above, the total degrees of freedom are the sum of the ones in each block.
\end{itemize}
Note that in the AR(1) model, the order of the effects in $\bm{r}_i$ is important.
An example in which we may want to leverage this fact is the case of many waves, when the random parameter for the effect at given wave $m$, depends on its values in the previous waves.
Another example is a model with a random effect in an ordered set of networks, which is encoded as a set of graphs in which the tie $(ij)$ can become one only if the same tie is already 1 in the ``lower" graphs.
In this case the random parameter in a given level, can depend on its values in the lower levels.

The models that are used as default are often \emph{diagonal}, which in the examples above are unrestricted diagonal, scalar, or block-diagonal with scalar blocks.
The reason is that they are advantageous computationally, but also statistically.
The methods for model evaluation that will be described in Section~\ref{section:ModelEvaluationComputationStandardErrors} for the case with one random effect, are easily generalizable to deal with multiple random effects, when the variance is diagonal.
Otherwise the generalization can be quite difficult, and possibly specific for a given model of the variance.

\section{Estimation method}
\label{sec:estimation}

In this section we describe the joint estimation of the parameters $\bm{\beta},\bm{\Sigma}\in\mathbb{R}^p\times \mathbb{S}^+_q$, where $\mathbb{S}^+_q$ is the set of $q$ dimensional positive definite matrices, by means of an extension of the \emph{simulated method of moments} (MoM).
For the standard stochastic actor oriented model, the MoM is discussed in \cite{snijders2001statistical} and \cite{snijders2007no}. For simplicity of notation, we will focus in this section on \emph{conditional} estimation of the rate parameters \citep[see Section 4.2]{snijders2001statistical}.
However, in Section~\ref{section:EstimationKapferer} the models will be estimated with unconditional estimation.
Other estimators developed for the SAOM include a Bayesian estimator \citep{KoskinenSnijders07}, a \emph{simulated maximum likelihood} estimator \citep{snijders2010maximum}, and a \emph{simulated generalized method of moments estimator} \citep{ASS2015,ASS2019}.
We base our estimation method on the MoM because likelihood-based algorithms require computation times which are more than 10 times longer.
Details about the algorithms are in \citet{SieAlg}.

\subsection{Simulated method of moments in the SAOM}

We first consider the SAOM without random effects.
The $p$ dimensional \emph{moment equation} for $\bm{\beta}$ is
\begin{equation}
E\bigg(\sum_{i=1}^N
\bm{s}_{i}(\bm{X}(t_2))\,\Big|\,\bm{X}(t_1)=\bm{x}(t_1),\bm{\beta}\bigg)=
\sum_{i=1}^N \bm{s}_{i}(\bm{x}(t_2)),
\label{equation:momentEquationBetaSAOM}
\end{equation}
the right side of the equation is called \emph{target}. 
The simulated method of moments procedure for solving this equation is based on simulating the generative Markov process from $t_1$ to $t_2$ for a given parameter $\bm{\beta}$, starting from the network at time $t_1$. 
The difference between the statistics of the simulated and observed networks at time $t_2$, is used to update the parameter. 
The network at time $t_2$ simulated for parameter $\bm{\beta}$ is denoted by $\hat{\bm{x}}(\bm{\beta})$, and its statistics are $\hat{\bm{s}}(\bm{\beta}) = \sum_i\bm{s}_i(\hat{\bm{x}}(\bm{\beta}))$.
The observed network at time $t_2$ and the target (observed statistics) are denoted by $\bm{x}$ and $\bm{s} = \sum_i\bm{s}_i(\bm{x})$, respectively.
The estimation algorithm is iterative, so a chain $(\bm{\beta}_{\texttt{t}})_{\texttt{t}}$ of parameters is generated via the  procedure
\begin{equation}
\bm{\beta}_{\texttt{t+1}} \xleftarrow{} \bm{\beta}_{\texttt{t}} -
\epsilon \bm{D}^{-1}_0\big(
\hat{\bm{s}}(\bm{\beta}_{\texttt{t}}) -
{\bm{s}}\big),
\label{equation:RobbinsMonroStepbeta}
\end{equation}
where $\epsilon$ is a positive \emph{gain factor}, and $\bm{D}^{-1}_0$ is an invertible \emph{preconditioning matrix}, computed at the beginning of the algorithm to compensate for the different magnitude and sensitivity of the components of $\hat{\bm{s}}(\bm{\beta})$  with respect to variations of $\bm{\beta}$.
This update rule is known as the \emph{Robbins-Monro algorithm} \citep{robbins1951stochastic}.

The parameter that solves (\ref{equation:momentEquationBetaSAOM}) is approximated by the tail average of the chain of parameters $(\bm{\beta}_{\texttt{t}})_{\texttt{t}}$.
This method is based on \cite{polyak1992acceleration}.
A small gain factor in  (\ref{equation:RobbinsMonroStepbeta}) ensures that the solution is approximated more accurately, assuming that the number of iterations is large enough so that the process has converged stochastically.
However, a small gain factor increases the auto-correlations between the elements in the chain $(\bm{\beta}_{\texttt{t}})_{\texttt{t}}$ so that more iterations are required to obtain a ``good" approximation.
In the algorithm used in the \emph{RSiena} package, the problem is solved with multiple \emph{sub-phases}, each with constant gain factor, and a ``provisional" estimate is used as starting point for the next sub-phase, after that the number of iterations, and the gain factor, for next sub-phase are increased, and decreased (by multiplying it by a reduction factor, which is 0.5 by default), respectively.
The last sub-phase produces the estimator $\hat{\bm{\beta}}$.
The current practice is to run the estimation algorithm with a given number of sub-phases (default 4) and a given number of runs per sub-phase.
The default number of iterations used in each sub-phase is described in page 393 of \citet{snijders2001statistical}.

After the estimator is computed, it is checked whether it is a good approximation of the solution of the moment equation (\ref{equation:momentEquationBetaSAOM}).
This is done by simulating the process multiple times with parameter $\hat{\bm{\beta}}$, and checking whether the simulated and observed statistics are equal on average, otherwise more sub-phases are needed.
The criterion of convergence is described in detail in \citet[Section 3.2]{SieAlg}.
These simulations are also used to approximate the variance of the estimator, as discussed in Section \ref{section:ModelEvaluationComputationStandardErrors}.

\subsection{Estimation with random effects}

In the stochastic actor oriented model with a single random effect, the moment equation for $\bm{\beta}$ remains the same as (\ref{equation:momentEquationBetaSAOM}), but a further equation is required for estimating the variance $\sigma^2$ of the random parameters. 
For the model with random out-degree the moment equations are
\begin{equation}
\begin{split}
&E\bigg(\sum_{i=1}^N
\bm{s}_{i}(\bm{X}(t_2))\,\Big|\,\bm{x}(t_1),\bm{\beta},\sigma^2\bigg)=
\sum_{i=1}^N \bm{s}_{i}(\bm{x}(t_2)), \\
&E \bigg( \sum_{i=1}^N
\big(s_{i1}(\bm{X}(t_2)) - \bar{s}_{1}(\bm{X}(t_2))\big)^2\,\Big|\,\bm{x}(t_1),\bm{\beta},\sigma^2 \bigg) =
\sum_{i=1}^N\big(s_{i1}(\bm{x}(t_2)) - \bar{s}_{1}(\bm{x}(t_2))\big)^2,
\end{split}
\label{equation:momentEquationsRandomOutDegree}
\end{equation}
that are to be solved for $\bm{\beta},\sigma^2$ simultaneously.
The first component is the $i$-th out-degree ${s}_{i1}(\bm{x}) = x_{i+}$, and $\bar{s}_1(\bm{x})$ is their average.

In the simulations for the parameter updates, model (\ref{equation:RandomOutDegUtilityFunction}) is used and random effects $b_{i}$ are independently sampled from $\mathcal{N}(0,\sigma^2_{\texttt{t}})$ for each parameter update step $\texttt{t}$.
The algorithm for $\bm{\beta}$ is the same as (\ref{equation:RobbinsMonroStepbeta}), replacing 
$\hat{\bm{s}}(\bm{\beta}_{\texttt{t}})$ with $\hat{\bm{s}}(\bm{\beta}_{\texttt{t}}, \bm{b}({\sigma}^2_{\texttt{t}}))=\bm{s}(\hat{\bm{x}}(\bm{\beta}_{\texttt{t}}, \bm{b}({\sigma}^2_{\texttt{t}}))$.
The update step for $\sigma^2$ is
\begin{equation}
\sigma^2_{\texttt{t+1}} \xleftarrow{}
    \text{max}\big( \sigma^2_{\texttt{t}} - \zeta \big(\hat{w}( \bm{\beta}_{\texttt{t}}, \bm{b}(\sigma^2_{\texttt{t}})) - w\big),\, \sigma^2_{\text{min}}\big), 
\label{equation:RobbinsMonrosigma2twoalgorithms}
\end{equation}
where $\zeta$ is a positive gain factor, the observed statistic for the variance is denoted by $w=\sum_{i=1}^N\big(s_{i1}(\bm{x}) - \bar{s}_{1}(\bm{x})\big)^2$, and the simulated statistic $\hat{w}( \bm{\beta}_{\texttt{t}}, \bm{b}(\sigma^2_{\texttt{t}}))$ is defined in a similar way by replacing the observed network $\bm{x}$ with the simulated network $\hat{\bm{x}}( \bm{\beta}_{\texttt{t}}, \bm{b}(\sigma^2_{\texttt{t}}))$.
The map $\tilde\sigma^2\mapsto \sigma^2= \max(\tilde\sigma^2, \sigma^2_{\text{min}})$ is used to force the updated variance, here denoted with $\tilde\sigma^2$, to be positive, where $\sigma^2_{\text{min}}$ is a very small positive value, for example $10^{-4}$.

The generalization to multiple random effects is straightforward.
In the evaluation function (\ref{equation:linearRandomUtilityFunction}) the number of random effects $q$ is larger than one.
Assume first that the variance is unconstrained, so that $\bm{\Sigma}\in\mathbb{S}_q^+$ has $q(q+1)/2$ parameters that are free to vary.
The moment equations are
\begin{equation}
\begin{split}
&E\bigg(\sum_{i=1}^N
\bm{s}_{i}(\bm{X}(t_2))\,\Big|\,\bm{x}(t_1),\bm{\beta},\bm{\Sigma}\bigg)=
\sum_{i=1}^N \bm{s}_{i}(\bm{x}(t_2)), \\
&E_{} \bigg( \sum_{i=1}^N
(\bm{r}_{i}(\bm{X}) - \bar{\bm{r}}(\bm{X}))(\bm{r}_{i}(\bm{X}) - \bar{\bm{r}}(\bm{X}))^\top \,\Big|\, \bm{x}(t_1),\bm{\beta},\bm{\Sigma} \bigg) =
\sum_{i=1}^N
(\bm{r}_{i}(\bm{x}) - \bar{\bm{r}}(\bm{x}))(\bm{r}_{i}(\bm{x}) - \bar{\bm{r}}(\bm{x}))^\top,
\end{split}
\label{equation:momentEquationsRandomEffects}
\end{equation}
where in the lower equation $\bm{X}$ and $\bm{x}$ stand for $\bm{X}(t_2)$ and $\bm{x}(t_2)$, respectively.
Note that the first equation is the same as the one in (\ref{equation:momentEquationsRandomOutDegree}).
Instead, the second equation is between two $q$ dimensional positive definite matrices.
The observed covariance matrix on the right hand side of the equation is denoted by $\bm{W}$.
If the simulated process is $\hat{\bm{x}}(\bm{\beta}, \bm{b}(\bm{\Sigma}))$, the simulated statistic is denoted succinctly as $\hat{\bm{W}}$, or more extensively as $\hat{\bm{W}}(\bm{\beta}, \bm{b}(\bm{\Sigma}))$ when the emphasis is on the parameters that have simulated the process.
The algorithm that generalizes (\ref{equation:RobbinsMonrosigma2twoalgorithms}) is
\begin{equation}
\bm{\Sigma}_{\texttt{t+1}}\xleftarrow{} \text{proj}\big(
\bm{\Sigma}_{\texttt{t}}
-\zeta \big(
 \hat{\bm{W}}(\bm{\beta}_{\texttt{t}}, \bm{b}(\bm{\Sigma}_{\texttt{t}})) - \bm{W}
\big), 
\sigma^2_\text{min}
\big),
\label{equation:projection}
\end{equation}
where the new value is projected (if necessary) to the space of positive definite matrices.
There are various ways to define the projector operator. A formal way is to use the method discussed in \cite{higham1988matrix}, where the matrix outside the parameter space is projected to the closest positive definite matrix (using the distance induced by a matrix norm).
Effectively, the method changes negative or zero eigenvalues to $\sigma^2_\text{min}$, before recomputing $\bm{\Sigma}_{\texttt{t+1}}$.

For the model with unconstrained diagonal variance: $\bm{\Sigma} = \mathrm{diag}(\sigma_h^2)$, the second equation of (\ref{equation:momentEquationsRandomEffects}) is replaced by the set of $q$ equations
\begin{equation}
E_{} \bigg( \sum_{i=1}^N
({r}_{ih}(\bm{X}) - \bar{{r}}_h(\bm{X}))^2 \,\Big|\, \bm{x}(t_1),\bm{\beta},\bm{\Sigma} \bigg) =
\sum_{i=1}^N
({r}_{ih}(\bm{x}) - \bar{{r}}_h(\bm{x}))^2,
\end{equation}
for $h\in\{1,...,q\}$, the update step is
\begin{equation}
\bm{\Sigma}_{\texttt{t+1}}\xleftarrow{} \text{proj}\big(
\bm{\Sigma}_{\texttt{t}}
-\zeta \big(
 \text{diag}(\hat{\bm{W}}(\bm{\beta}_{\texttt{t}}, \bm{b}(\bm{\Sigma}_{\texttt{t}}))) - \text{diag}(\bm{W})
\big), 
\sigma^2_\text{min}
\big),
\end{equation}
and the projection is much simpler.

When estimating a model, the usual assumption of \emph{identifiability} of the parameters is made. For estimation by the method of moments, this means that the moment equation (\ref{equation:momentEquationsRandomEffects}) has a unique solution with probability 1. This is guaranteed if the derivative of each expected statistic with respect to its associated parameter is monotonic, when all other parameters are kept constant. Most model specifications lead to mathematically identifiable models. However, it is possible that there is not enough information in a given dataset to estimate some effects together, due to a high level of multi-collinearity. In this case, some components of the process $(\bm{\beta}_{\texttt{t}}, \bm{\Sigma}_{\texttt{t}})_{\texttt{t}}$ fail to converge. One can either penalize the effects (e.g. ridge) or remove on of the ``problematic" covariates. This is in principle no different from a standard linear model. Therefore, failures of identifiability are usually identified during estimation. In Appendix~\ref{sec:identifiability}, the chains of parameters and simulated statistics for some effects of two models discussed in Section~\ref{section:EstimationKapferer} are shown. Although both models are theoretically identifiable, in one case there is not enough information in the data to estimate all model parameters accurately. 

Various random network models, specifically ERGMs, have an issue in estimation called \emph{degeneracy}, where the model, based on specified parameters, produces either nearly all-empty or nearly all-full networks, failing to capture the diversity of observed social structures \citep{handcock2003statistical}.
If the transitive triplets effect, that is usually responsible for degeneracy in ERGMs, is included, then for the usual rates of change (up to 20 transitions per actor from $t_1$ to $t_2$) and for the usual number of actors (less than $500$), the probability that the network explodes to an extremely high-degree network at $t_2$ is negligibly small. 
Therefore, in practice, transitive triplets and other effects that can cause degeneracy in various network models can be used in SAOM without problems.

\section{Model evaluation}
\label{section:ModelEvaluationComputationStandardErrors}

Model evaluation procedures for the SAOM with random out-degree and conditional estimation of the rate parameter, are introduced in Sections \ref{sec:ScoreTypeTestForOverdispersion} and \ref{sec:standardErrorsForSAOMWithRandomOutDegree}. In Section \ref{sec:generalizationsOfModelEvaluation},
some generalizations to more complex models are briefly discussed.
The model evaluation procedure presented here is a generalization of the test proposed by \citet{schweinberger2012statistical}, which requires the Monte-Carlo approximation of the derivative of the expected statistics with respect to the parameters, developed in \citet{schweinberger2007markov}.

\subsection{Score-type test for overdispersion}
\label{sec:ScoreTypeTestForOverdispersion}

The combination of the estimation functions for  $\bm{\beta}$ and $\sigma^2$
is denoted by $\bm{g}(\bm{x})=(\bm{s}(\bm{x}), w(\bm{x}))\in\mathbb{R}^{p+1}$.
The null hypothesis of the absence of overdispersion in the out-degree is
\begin{equation}
    H_0:\sigma^2 = 0,
\label{eq:H0}
\end{equation}
and can be tested against $H_1:\sigma^2 > 0$ with a \emph{score-type test} that will be soon described.
In linear random effects models, the score test is also known to be a good test for testing variance components; see \cite{berkhof2001variance}.

Expected value and variance of the statistics, which are
\begin{equation}
\bm{m}(\bm{\beta})=E_{\bm{\beta}}(\bm{g}(\bm{X})), \ \ \
\bm{V}(\bm{\beta}) = E_{\bm{\beta}}\big((\bm{g}(\bm{X})-\bm{m}(\bm{\beta}))(\bm{g}(\bm{X})-\bm{m}(\bm{\beta}))^\top\big),
\end{equation}
are approximated with Monte-Carlo integration as
\begin{equation}
\bar{\bm{m}}(\bm{\beta}) = \frac{1}{\texttt{T}}\sum_{\texttt{t=1}}^{\texttt{T}}\hat{\bm{g}}_{\texttt{t}}, \ \ \
\hat{\bm{V}}(\bm{\beta}) = \frac{1}{\texttt{T}}\sum_{\texttt{t=1}}^{\texttt{T}}(\hat{\bm{g}}_{\texttt{t}} - \bar{\bm{m}}(\bm{\beta}) )(\hat{\bm{g}}_{\texttt{t}} - \bar{\bm{m}}(\bm{\beta}) )^\top,
\label{eq:mbarbetaVhattheta}
\end{equation}
respectively,
where $\hat{\bm{g}}_{\texttt{t}}\in\mathbb{R}^{p+1}$ are
statistics simulated with parameters $\bm{\beta}\in\mathbb{R}^{p}$.
Developing a score-type test statistic with good power, requires however the ability to approximate the $(p+1)\times p$ dimensional matrix 
\begin{equation}
\bm{D}(\bm{\beta}) = 
\frac{\partial \bm{m}(\bm{\beta})}{\partial\bm{\beta}^\top}
=
\sum_{\bm{x}\in \bm{\mathcal{X}}}
\bm{g}(\bm{x})\Big(\frac{\partial}{\partial\bm{\beta}^\top}P(\bm{x}\mid \bm{\beta})\Big)
    \frac{P(\bm{x}\mid \bm{\beta})}{P(\bm{x}\mid \bm{\beta})}
=
E_{\bm{\beta}}
\Big(\bm{g}(\bm{X})\frac{\partial}{\partial\bm{\beta}^\top}\log P(\bm{X}\mid \bm{\beta})\Big),
\label{eq:Dbeta}
\end{equation}
where $P(\bm{x}\mid \bm{\beta})$ is the probability mass function of the network $\bm{x}$ according to parameter $\bm{\beta}$ \citep[Section 2]{rubinstein1989sensitivity}.
This derivative is approximated with Monte-Carlo integration as
\begin{equation}
\hat{\bm{D}}(\bm{\beta}) = \frac{1}{\texttt{T}}\sum_{\texttt{t=1}}^{\texttt{T}}\hat{\bm{g}}_{\texttt{t}}\hat{\bm{l}}^\top_{\texttt{t}}
\mbox{,~ or more stably as~  }
\hat{\bm{D}}(\bm{\beta}) = \frac{1}{\texttt{T}}\sum_{\texttt{t=1}}^{\texttt{T}} (\hat{\bm{g}}_{\texttt{t}} - \bm{g}) \hat{\bm{l}}^\top_{\texttt{t}},
\label{eq:hatJbetacomputedusingMC}
\end{equation}
where $\hat{\bm{g}}_{\texttt{t}} \in\mathbb{R}^{p+1}$  and $\hat{\bm{l}}^\top_{\texttt{t}}\in \mathbb{R}^{1\times p}$ are, respectively, the simulated statistics and the row-vector of contributions to the score function; and $\bm{g}$ is the vector of observed statistics. Note that $\hat{\bm{l}}^\top_{\texttt{t}}$ is the same row-vector that is computed for the model without random effects, because under the null hypothesis there are no random effects.
The only difference is that the variance of the out-degrees $\hat{w}_{\texttt{t}}$, which is the last component of $\hat{\bm{g}}_{\texttt{t}}$, has to be also computed, even though this statistic is not used to estimate the parameters.

The derivation of the test statistic that will be soon described is based on \citeauthor{neyman1959optimal}'s
(\citeyear{neyman1959optimal}) \emph{orthogonalization method}, which has been then developed in \citet{basawa1985neyman, basawa1991generalized}.
The orthogonalization procedure is a linear transformation that produces the test statistic from $\bm{g}(\bm{x})$, to obtain a test with high statistical power isolating the effect of $\sigma^2$ on the test statistic.
The procedure is presented here under the simplifying assumption of normality of the test statistic, however the \emph{p}-value $\hat{\alpha}$ derived below does not require normality, as the only important assumption is the finiteness of the first two moments $\bm{m}(\bm{\beta})$ and $\bm{V}(\bm{\beta})$ of the test statistic $\bm{g}(\bm{X})$.

Assume that
\begin{equation}
\bm{g}(\bm{X})\sim \mathcal{N}(\bm{m}(\bm{\beta}), \bm{V}(\bm{\beta})),
\label{equation:asymptoticDiscributionSufficientStatisticsH0nooverdispoersion}
\end{equation}
where the vector $\bm{m}(\bm\beta)$ and the matrices $\bm{D}(\bm\beta)$ and $\bm{V}(\bm\beta)$ are decomposed, in correspondence to $\bm{g}(\bm{x})=(\bm{s}(\bm{x}), w(\bm{x}))$, as
\begin{equation}
\bm{m}(\bm\beta)=\begin{pmatrix}
\bm{m}_1(\bm\beta) \\ {m}_2(\bm\beta)
\end{pmatrix}, \ \bm{D}(\bm\beta)=\begin{pmatrix}
\bm{D}_1(\bm\beta) \\ \bm{d}_2(\bm\beta)^\top
\end{pmatrix}, \
\bm{V}(\bm\beta)=\begin{pmatrix}
\bm{V}_{11}(\bm\beta) &\bm{v}_{12}(\bm\beta) \\ \bm{v}_{12}(\bm\beta)^\top &{v}_{22}(\bm\beta)
\end{pmatrix}.
\end{equation}
The component of $w(\bm{X})$ that is uncorrelated with $\bm{s}(\bm{X})$ is
\begin{equation}
{Y}(\bm\beta) = w(\bm{X})-\bm{\gamma}(\bm{\beta})^\top\bm{s}(\bm{X})\sim \mathcal{N}({m}_2(\bm\beta)-\bm{\gamma}(\bm{\beta})^\top\bm{m}_1(\bm\beta), \xi(\bm\beta)),
\end{equation}
where
\begin{equation}
\bm{\gamma}(\bm\beta)^\top=\bm{d}_2(\bm\beta)^\top\bm{D}_1(\bm\beta)^{-1}, \ \
\xi(\bm{\beta}) = v_{22}(\bm{\beta}) -
2\bm{\gamma}(\bm\beta)^\top\bm{v}_{12}(\bm{\beta}) + \bm{\gamma}(\bm\beta)^\top\bm{V}_{11}(\bm{\beta})\bm{\gamma}(\bm\beta).
\label{equation:GammaAndXiOfBeta}
\end{equation}
If $\bm{s}$ and $w$ are the observed statistics, the estimated parameter $\hat{\bm\beta}$ solves the moment equation $\bm{m}_1(\bm{\beta})=\bm{s}$, but under the null hypothesis also ${m}_2(\bm{\beta})=w$, as $\sigma^2=0$.
With equation (\ref{equation:asymptoticDiscributionSufficientStatisticsH0nooverdispoersion}) this shows that
\begin{equation}
\xi(\hat{\bm{\beta}})^{-1/2}({Y}(\hat{\bm\beta}) - {y}(\hat{\bm\beta}))\sim \mathcal{N}(0,1)
\label{equation:distributionstatistic2}
\end{equation}
under $H_0$, where ${y}(\hat{\bm\beta})= w-\bm{\gamma}(\hat{\bm{\beta}})^\top\bm{s}$.

The \emph{score-type test statistic}
$z$ derived from \cite{schweinberger2012statistical}, and the associated \emph{p-value} $\tilde{\alpha}$ are
\begin{equation}
z(\hat{\bm\beta}) = \hat\xi(\hat{\bm{\beta}})^{-1/2}({y}(\hat{\bm\beta}) - \bar{y}(\hat{\bm\beta})), \ \ \ \tilde{\alpha} = 1 - \Phi(z(\hat{\bm\beta})),
\label{equation:chisqtest2}
\end{equation}
where $z\mapsto \Phi(z)$ is the cumulative distribution function of the standard normal, $\bar{{y}}(\hat{\bm{\beta}}) = \bar{{m}}_2 - \hat{\bm{\gamma}}(\hat{\bm{\beta}})^\top\bar{\bm{m}}_1$
is the average of the simulated statistics, ${{y}}(\hat{\bm{\beta}})$ and $\hat{\xi}(\hat{\bm{\beta}})$ are computed with $\bm{\gamma}(\hat{\bm{\beta}})^\top$ replaced by $\hat{\bm{\gamma}}(\hat{\bm{\beta}})^\top$, substituting $\hat{\bm{D}}(\hat{\bm{\beta}})$ for ${\bm{D}}(\hat{\bm{\beta}})$, and $\hat{\bm{V}}(\hat{\bm{\beta}})$ for ${\bm{V}}(\hat{\bm{\beta}})$ in  equation (\ref{equation:GammaAndXiOfBeta}).
The computation of $\tilde{\alpha}$ relies on the normality in  (\ref{equation:distributionstatistic2}),  derived from the asymptotic normality in  (\ref{equation:asymptoticDiscributionSufficientStatisticsH0nooverdispoersion}).
The generalization proposed here relaxes the normality assumption, and the empirical distribution in the simulations is used to compute the $p$-value.
From the statistics $(\hat{\bm{g}}_{\texttt{t}})_{\texttt{t}}$ simulated in Phase 3 with parameter $\hat{\bm{\beta}}$, the matrices $\hat{\bm{V}}$ and $\hat{\bm{D}}$ are computed, and then used to compute the simulated test statistics $(\hat{{y}}_{\texttt{t}} = \hat{w}_{\texttt{t}}-\hat{\bm{\gamma}}^\top\hat{\bm{s}}_{\texttt{t}})_{\texttt{t}}$, and their mean $\bar{{y}} = \bar{{m}}_2 - \hat{\bm{\gamma}}^\top\bar{\bm{m}}_1$.
The dependence of these quantities on the estimate $\hat{\bm{\beta}}$ is implicit, to simplify the notation.
Then the \emph{empirical p-value} for testing (\ref{eq:H0}) is defined as
\begin{equation}
\hat\alpha = \frac{1}{\texttt{T}}\sum_{\texttt{t=1}}^{\texttt{T}}
I(\hat{{y}}_{\texttt{t}} \ge {{y}}) = \frac{1}{\texttt{T}}\sum_{\texttt{t=1}}^{\texttt{T}}
I(\hat{{z}}_{\texttt{t}} \ge {{z}}),
\label{eq:empiricalp}
\end{equation}
where $\hat{{z}}_{\texttt{t}} = \hat\xi^{-1/2}(\hat{y}_{\texttt{t}} - \bar{y})$, and $z=z(\hat{\bm{\beta}})$ is defined in (\ref{equation:chisqtest2}).

\subsection{Standard errors for SAOM with random out-degree}
\label{sec:standardErrorsForSAOMWithRandomOutDegree}

For computing the standard error of the estimates  $(\hat{\bm{\beta}}, \hat{\sigma}^2)$, the quantities $\bar{\bm{m}}(\hat{\bm{\beta}}, \hat{\sigma}^2)$ and $\hat{\bm{V}}(\hat{\bm{\beta}}, \hat{\sigma}^2)$ are approximated in the same way as in equation (\ref{eq:mbarbetaVhattheta}), but the simulated statistics $\hat{\bm{g}}_{\texttt{t}}$ used to compute these quantities are simulated with parameters $\hat{\bm{\beta}}$ and $\hat\sigma^2$.
However, the procedure for calculating $\hat{\bm{D}}(\hat{\bm{\beta}}, \hat{\sigma}^2)$ is slightly more challenging than the one in the preceding section.
Using the \emph{delta method} \citep[Theorem 5.13]{wasserman2004all}, the asymptotic covariance of the method of moment estimator $(\hat{\bm{\beta}}, \hat{\sigma}^2)$ is approximated by
\begin{equation}
\hat{\bm{C}}(\hat{\bm{\beta}}, \hat{\sigma}^2) =
\hat{\bm{D}}(\hat{\bm{\beta}}, \hat{\sigma}^2)^{-1}
\hat{\bm{V}}(\hat{\bm{\beta}}, \hat{\sigma}^2)
\hat{\bm{D}}(\hat{\bm{\beta}}, \hat{\sigma}^2)^{-\top},
\label{eq:ChatBetahatsigma2hat}
\end{equation}
see \citet{snijders2001statistical}, \citet[Theorem 9.6]{wasserman2004all}, $\hat{\bm{D}}^{-\top}$ denotes the inverse of the transpose of $\hat{\bm{D}}$.
The standard errors are the square roots of the diagonal elements of this matrix.

The $(p+1)\times(p+1)$ dimensional matrix $\bm{D}(\bm{\beta},\sigma^2)$ that generalizes (\ref{eq:Dbeta}) is
\begin{equation}
\begin{split}
 \bm{D}(\bm{\beta},\sigma^2) =
\frac{\partial \bm{m}(\bm{\beta},\sigma^2)}{\partial(\bm{\beta}, \sigma^2)^\top}
&=
\sum_{\bm{x}\in \bm{\mathcal{X}}}
\bm{g}(\bm{x})\Big(\frac{\partial}{\partial(\bm{\beta}, \sigma^2)^\top}\log P(\bm{x}\mid \bm{\beta},\sigma^2)\Big)
.
\end{split}
\label{eq:Jbetasigma2DefinitionDoubleIntegral}
\end{equation}
With $\bm{\eta} = \bm{h}(\bm{\beta},\sigma^2, \bm{u}) = (\bm{\beta},\bm{b})$, $\bm{b}=\sigma\cdot \bm{u}$, $\bm{u}\sim \mathcal{N}(\bm{0},\bm{I})$, the row-vector $\bm{l}^\top = (\bm{l}^\top_{\bm{\beta}} \ \ l_{\sigma^2})\in\mathbb{R}^{1\times(p+1)}$ of contributions to the score functions can be written using the chain rule in equation (\ref{equation:JacobianChainRule}) from Appendix \ref{section:CommDiagramsChainRule} as
\begin{equation}
\begin{split}
\bm{l}^\top
&
=
\frac{\partial}{\partial(\bm{\beta}, \sigma^2)^\top}\log P(\bm{x}\mid \bm{\beta},\sigma^2)
=
\frac{\partial}{\partial(\bm{\beta}, \sigma^2)^\top}\log \big( P(\bm{x}\mid \bm{\beta},\sigma^2, \bm{u})\phi(\bm{u})\big)
=
\\
&
=
\frac{\partial}{\partial\bm{\eta}^\top} \log P(\bm{x}\mid \bm{\eta}) \cdot 
\frac{\partial \bm{h}(\bm{\beta},\sigma^2, \bm{u})}{\partial (\bm{\beta}, \sigma^2)^\top}
=
\bm{l}^\top_{\bm{\eta}}
\begin{pmatrix}
\bm{I}_p &\bm{0}_p\\ \bm{O}_{N\times p} &\frac{1}{2\sigma^2}\bm{b}
\end{pmatrix}
= \Big(\bm{l}^\top_{\bm{\beta}} \ \ \frac{1}{2\sigma^2}\bm{l}^\top_{\bm{b}}\bm{b}\Big),
\end{split}    
\end{equation}
where $\phi(\bm{u})$ is the density of the multivariate standard normal in $\bm{u}$, which does not depend on $(\bm{\beta}, \sigma^2)$, and $\bm{l}^\top_{\bm{\eta}} = (\bm{l}^\top_{\bm{\beta}} \ \ \bm{l}^\top_{\bm{b}})\in\mathbb{R}^{1\times(p+N)}$.
The last component $l_{\sigma^2}=\frac{1}{2\sigma^2}\bm{l}^\top_{\bm{b}}\bm{b}$ of $\bm{l}^\top$, that has been computed using equation (\ref{equation:derivativeXwrtsigma2}) from Appendix \ref{section:derivativesGaussian}, is a random variable because $\bm{u}$, $\bm{b}$ and $\bm{\eta}$ are random vectors.

The matrix $\hat{\bm{D}}(\bm{\beta},\sigma^2)$ that approximates ${\bm{D}}(\bm{\beta},\sigma^2)$ is computed using the simulated statistics $\hat{\bm{g}}_{\texttt{t}} = (\hat{\bm{s}}_{\texttt{t}}, \hat{w}_{\texttt{t}})$ from the model with parameters $\bm{\beta}$ and $\sigma^2$.
In each simulation, the vector of random parameter $\bm{b}_{\texttt{t}}=\sigma \bm{u}_{\texttt{t}}$ is also simulated, and it is then used to compute the second component of the row-vector of contributions to the score function $\hat{\bm{l}}^\top_{\texttt{t}}$.
The generalization of the second matrix of equation (\ref{eq:hatJbetacomputedusingMC}), is then
\begin{equation}
\hat{\bm{D}}(\bm{\beta}, \sigma^2) = 
\frac{1}{\texttt{T}}\sum_{\texttt{t=1}}^{\texttt{T}}
\begin{pmatrix}
(\hat{\bm{s}}_{\texttt{t}} - \bm{s})\hat{\bm{l}}^\top_{\bm{\beta}\texttt{t}} &(\hat{\bm{s}}_{\texttt{t}} - \bm{s})\hat{l}_{\sigma^2\texttt{t}}\\
(\hat{w}_{\texttt{t}} - w)\hat{\bm{l}}^\top_{\bm{\beta}\texttt{t}} &(\hat{w}_{\texttt{t}} - w)\hat{l}_{\sigma^2\texttt{t}}
\end{pmatrix}
=
\frac{1}{\texttt{T}}\sum_{\texttt{t=1}}^{\texttt{T}}
\begin{pmatrix}
(\hat{\bm{s}}_{\texttt{t}} - \bm{s})\hat{\bm{l}}^\top_{\bm{\beta}\texttt{t}} &\frac{1}{2\sigma^2}(\hat{\bm{s}}_{\texttt{t}} - \bm{s})\hat{\bm{l}}^\top_{\bm{b}\texttt{t}}\bm{b}_{\texttt{t}}\\
(\hat{w}_{\texttt{t}} - w)\hat{\bm{l}}^\top_{\bm{\beta}\texttt{t}} &\frac{1}{2\sigma^2}(\hat{w}_{\texttt{t}} - w)\hat{\bm{l}}^\top_{\bm{b}\texttt{t}}\bm{b}_{\texttt{t}}
\end{pmatrix}.
\label{eq:Jhatbetasigma2asMCintegration}
\end{equation}

\subsection{Generalizations}
\label{sec:generalizationsOfModelEvaluation}

\paragraph{Different random effect.} As long as there is only one random effect in the model, all equations in Sections \ref{sec:ScoreTypeTestForOverdispersion} and \ref{sec:standardErrorsForSAOMWithRandomOutDegree} are the same even when the random effect is not the out-degree, using
$w(\bm{x})=\frac{1}{N}\sum_i(r_i(\bm{x})-\bar{r}(\bm{x}))^2$, the sample variance of the subgraph counts of the chosen random effect $\bm{r}$.

\paragraph{Unconditional estimation.} The generalization of the model evaluation procedure to the more common case in which the rate is estimated unconditionally, is also straightforward.
In this case $\bm{\beta}$, $p$ and $\bm{s}$, should be replaced in all equations above by $\bm{\theta}=(\bm{\lambda},\bm{\beta})$, $p^*=p_{\bm{\lambda}}+p$, and $\bm{s}^*=(\bm{s}_{\bm{\lambda}},\bm{s})$, respectively, where $\bm{\lambda}$ is the $p_{\bm{\lambda}}$ dimensional vector of rate parameters, and $\bm{s}_{\bm{\lambda}}(\bm{x})$ is the statistic used to estimate $\bm{\lambda}$.

\paragraph{Reparametrization.} Instead of computing the standard errors for $\bm{\beta}$ and $\sigma^2$, we might be interested in computing them for a different parametrization, e.g., for $\bm{\beta}$ and $\sigma$, where the second parameter is a standard deviation.
The derivative of $\sigma\mapsto \bm{b}=\sigma\bm{u}$ with respect to $\sigma$ is equal to $\bm{u}=\bm{b}/\sigma$.
Therefore, $\hat{\bm{D}}(\bm{\beta},\sigma)$ can be computed by replacing $(\bm{b}/2\sigma^2)$ with $\bm{u}$, and then $\sigma^2$ with $\sigma$, in all equations (\ref{eq:Jbetasigma2DefinitionDoubleIntegral} - \ref{eq:Jhatbetasigma2asMCintegration}).
Note that the matrix $\hat{\bm{D}}(\bm{\beta},\sigma)$ obtained by this ``replacement rule" is
\begin{equation}
\hat{\bm{D}}(\bm{\beta},\sigma) = \hat{\bm{D}}(\bm{\beta},\sigma^2)\bm{J}=\hat{\bm{D}}(\bm{\beta},\sigma^2)\begin{pmatrix}
\bm{I}_p &\bm{0}_p \\
\bm{0}_p^\top &2\sigma
\end{pmatrix} = \hat{\bm{D}}(\bm{\beta},\sigma^2) \left(\frac{\partial}{\partial(\bm{\beta}, \sigma)^\top}\Big[\begin{pmatrix}
\bm{\beta}\\ \sigma
\end{pmatrix}\mapsto \begin{pmatrix}
\bm{\beta}\\ \sigma^2
\end{pmatrix}\Big]\right),
\end{equation}
as $2\sigma\cdot(\bm{b}/2\sigma^2) = \bm{u}$.
Moreover $\hat{\bm{C}}(\hat{\bm{\beta}},\hat\sigma)$ computed using equation (\ref{eq:ChatBetahatsigma2hat}) with $\hat{\bm{D}}(\hat{\bm{\beta}},\hat\sigma)$, 
could have been derived directly with the delta method as $\hat{\bm{C}}(\hat{\bm{\beta}},\hat\sigma)=\bm{J}^{-1}\hat{\bm{C}}(\hat{\bm{\beta}},\hat\sigma^2)\bm{J}^{-\top}$, because $\bm{J}^{-1}$ is the derivative of $(\bm{\beta}, \sigma^2) \mapsto (\bm{\beta},\sigma)$ with respect to $(\bm{\beta}, \sigma^2)^\top$.
Therefore, the model evaluation procedure developed in the previous sections is coherent with the statistical theory of reparametrizations, which is important when the interest is on a particular interpretation of the estimated parameter.

\paragraph{Multiple random effects.} If there are more random effects, and they are independent, for example when $\bm{\Sigma}=\mathrm{diag}(\sigma_1^2, \sigma_2^2)$, the model evaluation procedure can be easily generalized for testing $H_0:\sigma_h^2=0$, or for computing the standard error of $(\bm{\beta},\sigma_1^2,\sigma_2^2)$.
In the latter case, the matrix $\hat{\bm{V}}(\bm{\beta}, \bm{\Sigma})$ is computed as in equation (\ref{eq:mbarbetaVhattheta}), but with the simulated statistic $\hat{\bm{g}}_{\texttt{t}}$ computed with $\bm{g}(\bm{x}) = (\bm{s}(\bm{x}), \mathrm{diag}(\bm{S}(\bm{x})))=(\bm{s}(\bm{x}), w_1(\bm{x}), w_2(\bm{x}))$.
The matrix $\hat{\bm{D}}(\bm{\beta}, \bm{\Sigma})$ can also be computed with a similar procedure, where $\bm{g}(\bm{x})\in\mathbb{R}^{p+2}$ is simulated, and the contribution to the score function is
\begin{equation}
\bm{l}^\top=(\bm{l}^\top_{\bm{\beta}} \ \ \bm{l}^\top_{\bm{\Sigma}}) = (\bm{l}^\top_{\bm{\beta}} \ \ {l}_{\sigma^2_1} \ \ {l}_{\sigma^2_2}) = \Big(\bm{l}^\top_{\bm{\beta}} \ \ \frac{1}{2\sigma_1^2}\bm{l}^\top_{\bm{b}_1}\bm{b}_1 \ \ \frac{1}{2\sigma_2^2}\bm{l}^\top_{\bm{b}_2}\bm{b}_2\Big)\in \mathbb{R}^{1\times(p+2)},
\end{equation}
where $\bm{b}_1$ and $\bm{b}_2$ are the vectors of random parameters for the two effects, so that $\hat{\bm{D}}(\bm{\beta}, \bm{\Sigma})$ can be computed with a formula similar to (\ref{eq:Jhatbetasigma2asMCintegration}).
For testing the null hypothesis of no overdispersion in one random effect, for example $H_0:\sigma^2_2=0$, the same procedure of Section \ref{sec:ScoreTypeTestForOverdispersion} is used, but with $\bm{l}^\top=(\bm{l}^\top_{\bm{\beta}} \ \ {l}_{\sigma^2_1})\in\mathbb{R}^{1\times (p+1)}$ and $\bm{g}(\bm{x}) =(\bm{s}(\bm{x}), w_1(\bm{x}), w_2(\bm{x})) \in \mathbb{R}^{p+2}$ used to approximate $\bm{D}(\bm{\beta},\sigma^2_1)\in \mathbb{R}^{(p+2)\times (p+1)}$, which can be used to transform the sufficient statistics into a test statistic with the orthogonalization procedure.
If the random effects are still independent, but some (variance) parameters are shared between different effects, it is still possible to extend easily the estimation and the model evaluation procedure to this case. The equations for $\bm{D}$ and $\bm{V}$ will depend on which parameters are shared, so we do not discuss them in detail.
The generalization of the model evaluation procedure is however much more difficult in the case in which the random effects are not independent.
The reason is that when $\bm\Sigma$ is diagonal, the maps $\bm\Sigma\mapsto \bm\Sigma^{1/2}$ and $\bm\Sigma^{1/2}\mapsto \bm{b}$ needed to correlate the random parameters, can be written simply as vector-to-vector functions $(0,\infty)^q\to (0,\infty)^q$ and $(0,\infty)^q\to\mathbb{R}^{qN}$, respectively.
By contrast, if some random effects are correlated, $\bm\Sigma\mapsto \bm\Sigma^{1/2}$ and $\bm\Sigma^{1/2}\mapsto \bm{b}$ must be treated as a matrix-to-matrix, and a matrix-to-vector function, respectively, complicating significantly the model evaluation procedure.

\paragraph{Composite hypotheses.} Finally, composite hypotheses on $\bm{\beta}$, $\bm{\Sigma}$, or a combination of the two are discussed.
Consider the hypothesis $H_0:\bm{\beta}_2=\bm{0}_{p_2}$ for $\bm{\beta}=(\bm{\beta}_1,\bm{\beta}_2)$, when there are random effects in the model, e.g., in the SAOM with random out-degree.
The simulated quantities under the null hypothesis are used to approximate the positive definite matrix $\bm{V}(\bm{\beta}_1,\sigma^2)\in\mathbb{S}_{p+1}^+$, and the Jacobian $\bm{D}(\bm{\beta}_1,\sigma^2)\in\mathbb{R}^{(p+1)\times (p-p_2+1)}$.
These quantities are used to compute
\begin{equation}
\begin{split}
\bm\Gamma=\bm{D}_2\bm{D}_1^{-1} \in \mathbb{R}^{p_2\times (p-p_2+1)}, \ \ \
\bm{\Xi} = \bm{V}_{22} -
(\bm{\Gamma}\bm{V}_{12}+\bm{V}_{12}^\top\bm{\Gamma}^\top) + \bm{\Gamma}\bm{V}_{11}\bm{\Gamma}^\top \in \mathbb{S}_{p_2}^+,
\end{split}
\end{equation}
where in order to keep the notation simple, the dependence of these quantities on $(\bm{\beta}_1,\sigma^2)$, the parameter under $H_0$, is implicit.
Therefore
$\bm{Z}=\bm{\Xi}^{-1/2}(\bm{Y}-\bm{y})\sim\mathcal{N}(\bm{0}_{p_2}, \bm{I}_{p_2})$ under the null hypothesis, for $\bm{Y}=\bm{g}_2(\bm{X})-\bm{\Gamma}\bm{g}_1(\bm{X})$, where $\bm{x}\mapsto \bm{g}_2(\bm{x})$ are the simulated statistics associated with the parameters that are $0$ under the null, and $\bm{y}$ is computed orthogonalizing the observed statistics.
The test statistic used in \cite{schweinberger2012statistical}, with its associated \emph{p}-value is
\begin{equation}
z^2 = (\bm{y}-\bar{\bm{y}})^\top \hat{\bm{\Xi}}^{-1}(\bm{y}-\bar{\bm{y}}), \ \ \ \tilde{\alpha}=1-\chi^2_{p_2}(z^2),
\label{eq:z2testsweinbgerger}
\end{equation}
where $\chi^2_{p_2}(z^2)$ is the cumulative distribution function of a Chi-squared distribution with $p_2$ degrees of freedom, evaluated at $z^2$, and $\bar{\bm{y}}$ is the average of the simulated test statistics $\hat{\bm{y}}_{\texttt{t}}$.
The non-parametric version of $\tilde{\alpha}$ that is based on the empirical distribution of the simulated test statistics, rather than assuming that their distribution is normal, is
\begin{equation}
\hat{\alpha}
=
\frac{1}{\texttt{T}}\sum_{\texttt{t=1}}^{\texttt{T}} I \big(
(\hat{\bm{y}}_{\texttt{t}}-\bar{\bm{y}})^\top \hat{\bm{\Xi}}^{-1} (\hat{\bm{y}}_{\texttt{t}}-\bar{\bm{y}}) \ge z^2
\big),
\end{equation}
where 
$z^2$ is defined in (\ref{eq:z2testsweinbgerger}).
The same theory can be used for hypotheses as $H_0:\sigma^2_h = 0$, for $h \in\ \{q_1+1,...,q\}$, in which the last $q_2=q-q_1$ random effects are not significant, assuming $\bm{\Sigma}$ is diagonal, or for hypotheses $H_0:\beta_k = 0, \sigma^2_h = 0$, for $k \in\ \{p_1+1,...,p\}$ and $h$ as before.
Therefore the theory developed in this section, generalizing  \cite{schweinberger2012statistical}, can be used to test most hypotheses in which a group of effects, fixed or random, is not included in the model under the null hypothesis, as long as the variance $\bm{\Sigma}$ of the random effects is diagonal.

\paragraph{Model selection.}
The disadvantage of score-type hypothesis testing discussed so far is that only nested models can be compared.
Model 1 $\mathcal{M}_1$ is determined by its estimated parameters $(\bm{\beta}_1,\bm{\Sigma}_1)\in\mathbb{R}^{p_1}\times (0,\infty)^{q_1}$ (assuming that the random effects are independent) and by the (observed) statistics $\bm{g}^1\in \mathbb{R}^{p_1+q_1}$ that are used to estimate the fixed parameters.
Similarly $\mathcal{M}_j=(\bm{\beta}_j,\bm{\Sigma}_j, \bm{g}^j)$ for all other models considered.
Consider the vector of observed statistic $\bm{g}^0\in \mathbb{R}^{p_0+q_0}$ that will be used to evaluate the different models.
All statistics that are present in at least one of the two models should be included in $\bm{g}^0$, but other statistics can be included as well.
Then $p_0\ge p_j$ and $q_0\ge q_j$ for all $j$.
The simulated statistics for model $j$ are $(\hat{\bm{g}}_{\texttt{t}}^{0(j)})_{\texttt{t}}$ and $\hat{\bm{V}}_{0(j)}$ is their variance.
For each model, a \emph{parameter selection criterion} is defined as
\begin{equation}
\mathrm{psc}(\mathcal{M}_j) = \mathrm{df}(N)\cdot \frac{1}{\texttt{T}}\sum_{\texttt{t=1}}^{\texttt{T}}\big( (\hat{\bm{g}}_{\texttt{t}}^{0(j)} - \bm{g}^0)^\top \hat{\bm{V}}_{0(j)}^{-1} (\hat{\bm{g}}_{\texttt{t}}^{0(j)} - \bm{g}^0) \big) - (p_0 - p_j + q_0 - q_j)\mathrm{pen}(\mathrm{df}(N)),
\label{eq:psc}
\end{equation}
where $\mathrm{df}(N)$ is an increasing function of the network size $N$, $\mathrm{pen}(\mathrm{df}(N))$ that is used to penalize model with more parameters, is equal to 2 and $\log(\mathrm{df}(N))$ for the \emph{AIC} and \emph{BIC} selection criteria, respectively.
The selected model is the one that minimizes the chosen parameter selection criterion.
This approach is an adaptation of \cite{andrews1999consistent}, in which a ``similar" \emph{moment selection criterion} was used to choose which set of statistics should be used to estimate with generalized method of moments the parameter of the model.
His approach differs from ours as the parameter is the same in all models that are evaluated, and the selection criterion is used to choose the statistics needed to estimate it.
We have modified the method so that the fixed set of statistics, that includes the ones used to estimate the models that are considered, is used to compare different parametrizations.

\paragraph{Goodness of fit.} 
The usual method to assess goodness of fit in the SAOM, that is very briefly described here, can be applied straightforwardly in models with random effects.
This method uses an auxiliary statistic, which contains features of the data that are not used to estimate the parameters of the model.
Denote with $\bm{a}=\bm{a}(\bm{x})$ the observed auxiliary statistic, with $\hat{\bm{a}}_{\texttt{t}}$ the simulated one at iteration $\texttt{t}$, with $\bar{\bm{a}}$ and $\hat{\bm{\Omega}}$ the mean and variance of $\hat{\bm{a}}_{\texttt{t}}$ across iterations.
The \emph{p}-value
\begin{equation}
\hat{\alpha}_{GOF} = 
\frac{1}{\texttt{T}}\sum_{\texttt{t=1}}^{\texttt{T}} I \big(
(\hat{\bm{a}}_{\texttt{t}}-\bar{\bm{a}})^\top \hat{\bm{\Omega}}^{-1} (\hat{\bm{a}}_{\texttt{t}}-\bar{\bm{a}}) \ge 
(\bm{a}-\bar{\bm{a}})^\top \hat{\bm{\Omega}}^{-1} (\bm{a}-\bar{\bm{a}})
\big),
\label{eq:pvalueGOF}
\end{equation}
can be used to test the null hypothesis in which the model which generated the observed data is the estimated
model.
Choosing a good auxiliary statistic can be difficult, various alternatives are discussed in \cite{lospinoso2019goodness}, in which this goodness of fit procedure is described in detail.

\section{Analysing social interactions in a tailor shop}
\label{section:EstimationKapferer}
In this section a dataset is analyzed, mainly with the purpose of illustrating the proposed model, its estimation, and its evaluation procedure.
The interpretation of some estimated parameters is described in detail, to show how the inclusion of the random out-degree effect can lead to different estimates of other parameters of interest.
The algorithm that has been used to estimate models with random effects is described in Appendix~\ref{sec:implementation dummy variables}, and the code used in this section is available in the GitHub repository \url{https://github.com/gceoldo/SAOM-with-random-effects}.

The main comparison in the analysis is between the model containing the out-degree activity, and the model containing the random out-degree.
These two effects, that have a different interpretation, can be used to explain the observed overdispersion in the out-degree distribution.
In particular, a positive parameter for the out-degree activity can increase the variability of the out-degree distribution, because individuals with many ties would like to form even more ties, leading to an out-degree distribution with an heavier tail.
We will also check whether the random out-degree can be used to replace the transitive triplets effect, as suggested by \citet{thiemichen2016bayesian} in the context of ERGMs, and on whether the random effect can replace the status covariate.

\subsection{Kapferer's tailor shop dataset}
\label{subsection:KapfererDataset}
Between June 1965 and January 1966 \cite{kapferer1972strategy} carried out a study of a tailor shop in Kabwe, Zambia, in a period when the workers were negotiating for higher wages. There were $39$ employees; networks between them were observed at two times, seven months apart.
The dataset used is available in the \emph{Ucinet} data repository \url{http://vlado.fmf.uni-lj.si/pub/networks/data/Ucinet/UciData.htm}.
We study the ``sociational'' (friendship, socioemotional) network.

\begin{figure}[tbh]
    \centering
    \includegraphics[scale=.65]{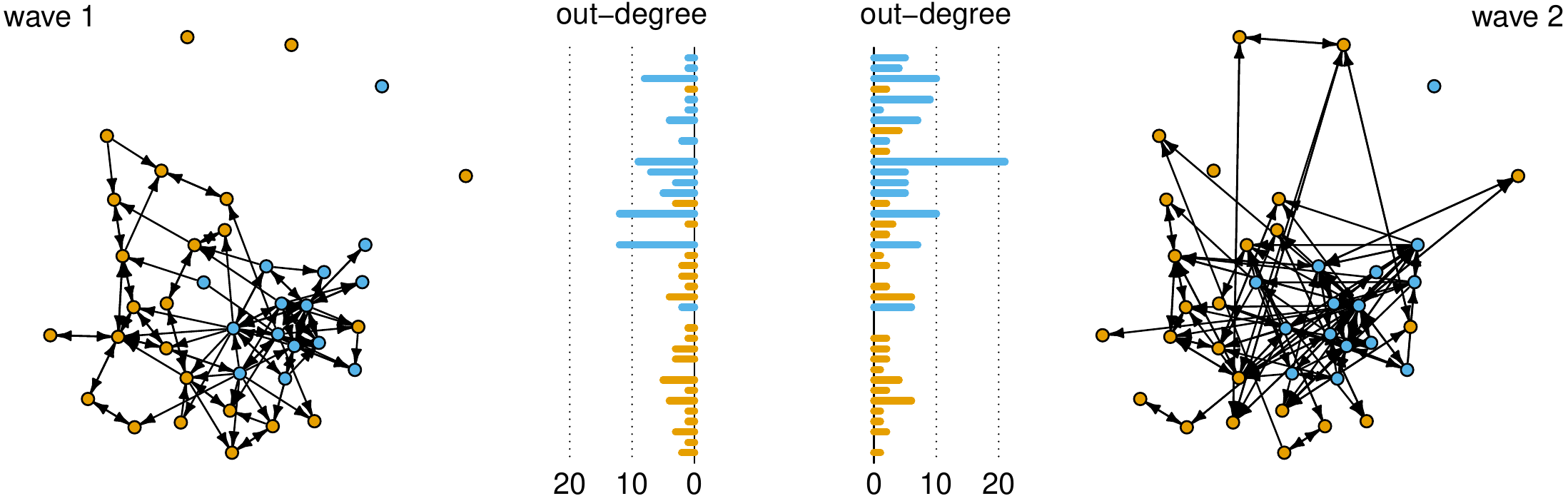}
    \caption{Kapferer data set with out-degrees of the nodes.}
    \label{figure:Kapferer}
\end{figure}

An important covariate is job status (low or high). We shall not consider the specific job of the individuals in our analysis.
Figure~\ref{figure:Kapferer} shows the two network snapshots with the out-degrees of the individuals. Each individual is in the same position in the two plots.
Status is indicated by color (light blue for high, orange for low).
Out-degrees in the first graph range from $0$ to $12$ with an average of $2.8$, whereas in the second graph they range from $0$ to $21$ with an average of $3.8$.

The dataset was described first in \cite{kapferer1972strategy}, and has been the subject of a variety of analyses.
\cite{carley1990group} modeled the stability of groups in the network.
The dataset was analysed in \cite{nowicki2001estimation} using a \emph{stochastic block model}.
In \cite{robins2004missing} the network at time~1 was used to check how the estimate of the parameters in an \emph{exponential random graph model} is influenced by various patterns of missing relations.
In \cite{caimo2014bergm} the data set was used to illustrate their implementation of Bayesian ERGMs.
We cannot compare the outcomes of our models with those of earlier analyses, because the objectives and models that we found in the literature are very dissimilar from our own.

\subsection{Definition estimated models}
\label{subsection:DefinitionEstimateModels}

The aim of the analysis is to discover what has been driving changes of ties in the network.
We are particularly interested in discovering to what extent triadic effects, such as transitivity and out-degree activity, as well as work status (low/high) might be confounded with heterogeneity in out-degree.
In order to answer these questions, we consider the following fixed effects,
\begin{itemize}
    \item \emph{basic rate}: rate of tie flips between the two waves;
    \item \emph{out-degree} (\emph{density}): number of ties starting from the focal actor $i$,  $s_{i1}(\bm{x})=\sum_jx_{ij}$;
    \item \emph{reciprocity}: number of reciprocated ties,  $s_{i2}(\bm{x})=\sum_jx_{ij}x_{ji}$;
    \item \emph{transitive triplets}: ordered pairs of actors $(jh)$ to both of whom $i$ is tied, while also $j$ is tied to $h$,  $s_{i3}(\bm{x})=\sum_{j,h}x_{ij}x_{ih}x_{hj}$;
    \item \emph{out-degree activity}: squared out-degree of the actor, $s_{i4}(\bm{x}) = (\sum_jx_{ij})^2$;
    \item \emph{status alter}:  sum of the covariate over all actors to whom $i$ has a tie, calculated as $s_{i5}(\bm{x}) = \sum_jx_{ij}v_j$, where $v_j$ is the binary covariate status, of actor $j$;
    \item \emph{status ego}: out-degree of actor $i$ weighted by its status, $s_{i6}(\bm{x}) = v_i\sum_jx_{ij}$;
    \item \emph{status similarity}: sum of centered similarity scores between $i$ and the other actors $j$ to whom he is tied, calculated as
    $$ s_{i7}(\bm{x}) = \sum_jx_{ij}\big(\mathrm{sim}_{ij}(\bm{v}) - \overline{\mathrm{sim}}(\bm{v})\big),$$
    where, for a binary covariate, $\mathrm{sim}_{ij}(\bm{v})= 1-|v_i-v_j|,$ and $\overline{\mathrm{sim}}(\bm{v})$ is their mean.
\end{itemize}
We consider four models with different selections of fixed effects, each of these four model is estimated with and without random out-degree.
The results are discussed in the next section, and summarized in Table~\ref{table:estimatesWithSE}.


\subsection{Estimation and interpretation of the results}
\label{subsection:estimation}
The models without random out-degree are estimated first using the function \texttt{siena07} of the package \emph{RSiena}.
Then the estimates are used as starting points for the models with random out-degree, using $\sigma^2_{min}=10^{-4}$ as starting value for $\sigma^2$.
Parameter update steps (\ref{equation:RobbinsMonroStepbeta}) and (\ref{equation:RobbinsMonrosigma2twoalgorithms}) were used, with in  (\ref{equation:RobbinsMonroStepbeta}) the replacement of $\bm{\beta}$ by $\bm{\theta}=(\lambda,\bm{\beta})$, because the rate parameter is estimated unconditionally.
To have a direct comparison between the algorithms, in all models the same gain factor $\epsilon$ is used, and in each sub-phase the gain factor is decreased in the same way.
This employs the default values of \texttt{siena07}, i.e., in the first sub-phase $\epsilon = 0.2$ and after the end of a sub-phase the gain factor is decreased by the factor~2.
For the models with random out-degree, the gain factor $\zeta$, in the first sub-phase, is set so that the magnitude of the updates of $\sigma^2$ is comparable with the updates of $\bm{\theta}$. 
The gain factor is also decreased by the factor~2 at the end of the sub-phase.
For each of the four models without random effects, the preconditioning matrix $\bm{D}_0^{-1}$ used for updating $\bm{\theta}$ is computed with the default procedure of \texttt{siena07}, and used also for the equivalent model with random effect.

Phase 2 consists of 2,100 iterations in total, divided in four sub-phases of 100, 100, 200 and 1,700 iterations.
The number of iterations used to compute the tail average used as starting point of the next sub-phase is 20, 40 and 80, for the first three sub-phases, respectively, and the last 1,500 iterations are averaged to compute the estimated parameter.
In Phase 3, 5,000 iterations are used in all models.
We have checked that this number of iterations is large enough to have stable approximations for the standard errors, and for the quantities computed for the model evaluation that are discussed in the next section.

The full model with random out-degree did not converge because of near collinearity, although the average simulated statistics for all effects were very near the observed ones. 
The near collinearity is caused by the presence among the estimation statistics of the sum of out-degrees (for the fixed out-degree parameter), the sum of squared outdegrees
(out-degree activity effect) and the variance of the out-degrees (random out-degree parameter). 
The latter is a deterministic, although not linear, function of the other two.
Moreover the sum of out-degrees for individual with high status (status-ego effect) was also in the model.
For this dataset there are different combinations of $\beta_{\mbox{\scriptsize out-degree activity}}$, $\beta_{\mbox{\scriptsize status ego}}$, and $\sigma^2$ that produce similar simulated statistics, so these three parameters cannot be estimated together with method of moments, making the full model with random out-degree ``almost non-identifiable", with this estimation method.
The chains of some effects in the optimization of this model are shown in Appendix~\ref{sec:identifiability}. 

\begin{figure}[tbh]
    \centering
    \includegraphics[scale=.49]{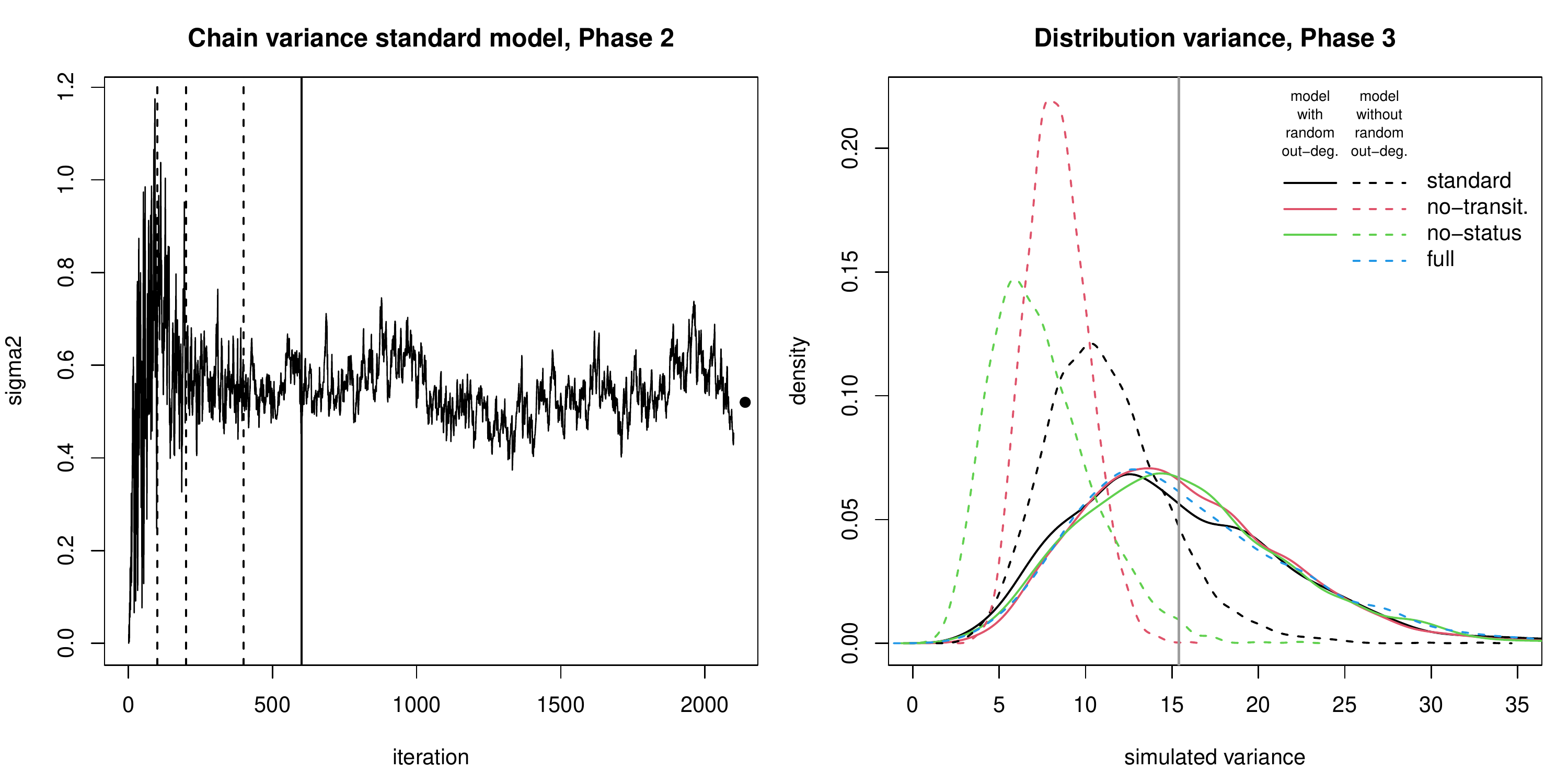}
    \caption{Left: chain of the variance parameter for the standard model with random out-degree. Right: distribution of the variance of the out-degrees computed from the simulations in Phase 3.}
    \label{figure:algorithm}
\end{figure}

In the left plot of Figure~\ref{figure:algorithm}, the chain of the variance parameter for the standard model with random out-degree is plotted.
The dashed vertical lines are the iterations where the gain factor is decreased, and the filled vertical line denotes the end of the burn-in period in which it is not assumed that the process has reached stationarity.
So all values on the right of the filled line are averaged to compute the estimated variance, which is the black dot on the right of the plot.
The right plot shows the distribution of the variances of the out-degrees of the networks for wave 2 as simulated in Phase 3, for all estimated models.
Filled and dashed lines are for the models with and without random out-degree, respectively.
The colours denote the models, they are black, red, green and blue for the standard, no-transitivity, no-status and full models, respectively.
The vertical grey line is the variance of the out-degrees in the observed network in wave 2.
The distributions for the models with random out-degree are all similar, and their average is the sample variance of the observed network, whereas for the models without random out-degree, only the full model explains the observed overdispersion.

Estimated parameters and standard errors are reported in Table~\ref{table:estimatesWithSE}.
The models without random out-degree have, in all cases, an estimated rate higher than the corresponding models with random out-degree.
This means that for the models with random out-degree, fewer steps are required, on average, to simulate a network with statistics that match the data.

Comparison of the parameter estimates for the fixed effects between the models with and without random out-degree shows that the absolute values for all parameters, except transitive triplets, are higher in the models with random out-degree.
To explain this, note that it was argued for logistic regression models by \citet[p. 68-69]{mood2010logistic} that,
since the error term is included implicitly in the non-linear link function, extending a given model by extra uncorrelated explanations will lead to approximately scaling the other parameters and their standard errors with a factor greater than~1.
The same holds for multinomial logistic models such as (\ref{equation:transitionProbabilityUtility}).
Therefore the main interpretation of the greater absolute values of the parameters and standard errors in the models with random out-degree is the increase in explanatory power of this model (reflected also by the lower estimated rate parameters).
The exception is the transitive triplets parameter, implying that the random out-degree effect partially explains the triangles in the model.
Special mention should be made here of the status ego effect, because ego effects of covariates are the deterministic, i.e., explained, analogues of the random out-degree effect.
The increase in the standard error of status ego is more than proportional with the other increases, but the effect stays significant.

\begin{table}[tbh]
\centering
\begin{tabular}{lcccc}
\specialrule{1.5pt}{1pt}{1pt}
\emph{Without random out-degree} & { \ \ standard \ \ } & no-transitivity & { \ \ no-status \ \ } & { \ \ \ \ \ full \ \ \ \ \ }  \\
\emph{effects / parameters}             & est. (s.e.)         & est. (s.e.)    & est. (s.e.)          & est. (s.e.) \\
\hline
basic rate                 & 21.39 (4.60)         & 19.05 (3.58)    & 16.68 (2.92)          & 22.94 (5.13) \\
out-degree (density)           & -2.66 (0.23)         & -2.52 (0.24)    & -2.08 (0.12)          & -2.97 (0.28) \\
reciprocity          & 3.26 (0.40)          & 3.37 (0.43)     & 2.23 (0.21)           & 3.34 (0.44)  \\
transitive triplets  & 0.19 (0.05)          & /               & 0.21 (0.04)           & 0.11 (0.05)  \\
out-degree activity              & /                    & /               & /                     & 0.04 (0.01) \\
status alter         & -1.15 (0.24)         & -0.98 (0.27)    & /                     & -1.06 (0.26) \\
status ego           & 1.45 (0.28)          & 1.77 (0.31)     & /                     & 1.24 (0.29)  \\
status similarity          & 0.30 (0.13)          & 0.42 (0.13)     & /                     & 0.40 (0.13)  \\
\hline
\emph{AIC par. selection criterion} &219.0 &724.9 &789.3 &85.7\\
\specialrule{1.5pt}{1pt}{1pt}
\end{tabular}

\vspace*{.5cm}
\begin{tabular}{lcccc}
\specialrule{1.5pt}{1pt}{1pt}
{\emph{With random out-degree} \ \ \ } & { \ \ standard \ \ } & no-transitivity & { \ \ no-status \ \ } & { \ \ \ \ \ full \ \ \ \ \ }  \\
\emph{effects / parameters }             & est. (s.e.)         & est. (s.e.)    & est. (s.e.)          & est. (s.e.) \\
\hline
basic rate                 & 17.30 (4.19)         & 15.78 (4.36)    & 10.86 (2.60)          &  \\
out-degree (density)           & -2.98 (0.32)         & -3.03 (0.40)    & -2.76 (0.48)          &  \\
reciprocity          & 3.64 (0.46)          & 3.87 (0.54)     & 3.16 (0.59)           &   \\
transitive triplets & 0.13 (0.07)          & /               & 0.16 (0.08)           &  \\
out-degree activity     & /                    & /               & /                     &  \\
status alter         & -1.17 (0.26)         & -1.06 (0.27)    & /                     &  \\
status ego           & 1.83 (0.43)          & 2.16 (0.56)     & /                     &   \\
status similarity         & 0.48 (0.18)          & 0.61 (0.19)     & /                     &   \\
\hdashline
variance random out-degree        & 0.52 (0.43)          & 0.92 (0.64)     & 1.92 (2.05)           &  \\
std.dev. random out-degree        & 0.72 (0.30)          & 0.96 (0.33)     & 1.39 (0.74)           & \\
\hline
\emph{AIC par. selection criterion} &86.7 &171.0 &333.3 & \\
\specialrule{1.5pt}{1pt}{1pt}
\end{tabular}
\caption{Estimated parameters with standard errors in brackets for the models with and without random out-degree. Full-transitivity model with random out-degree is missing, as the algorithm has not converged. For each estimated model the AIC parameter estimation criterion is plotted in the last row.} 
\label{table:estimatesWithSE}
\end{table}

Table~\ref{table:estimatesWithSE} also gives the estimates $\hat{\sigma}^2$ and its positive root $\hat{\sigma}$ of the variance and the standard deviation of the random out-degree effect, respectively, with their standard errors.
The estimated variance parameters are smaller for the models containing more effects, which suggests that omitting effects will lead to higher estimated residual heterogeneity.
The standard errors of $\hat{\sigma}$ are computed with the delta method from those of $\hat{\sigma}^2$.
Considering the ratios of these parameters to their standard errors highlights that these ratios should not be used in a conventional $t$-test for testing the null hypothesis that $\sigma^2=0$. The better strategy is to use a parameter selection criterion or a score-type test, as demonstrated in the next section.

\subsection{Model evaluation and comparison}
\label{subsec:hypothesisTests}
\begin{figure}[tbh]
    \centering
    \includegraphics[scale=.7]{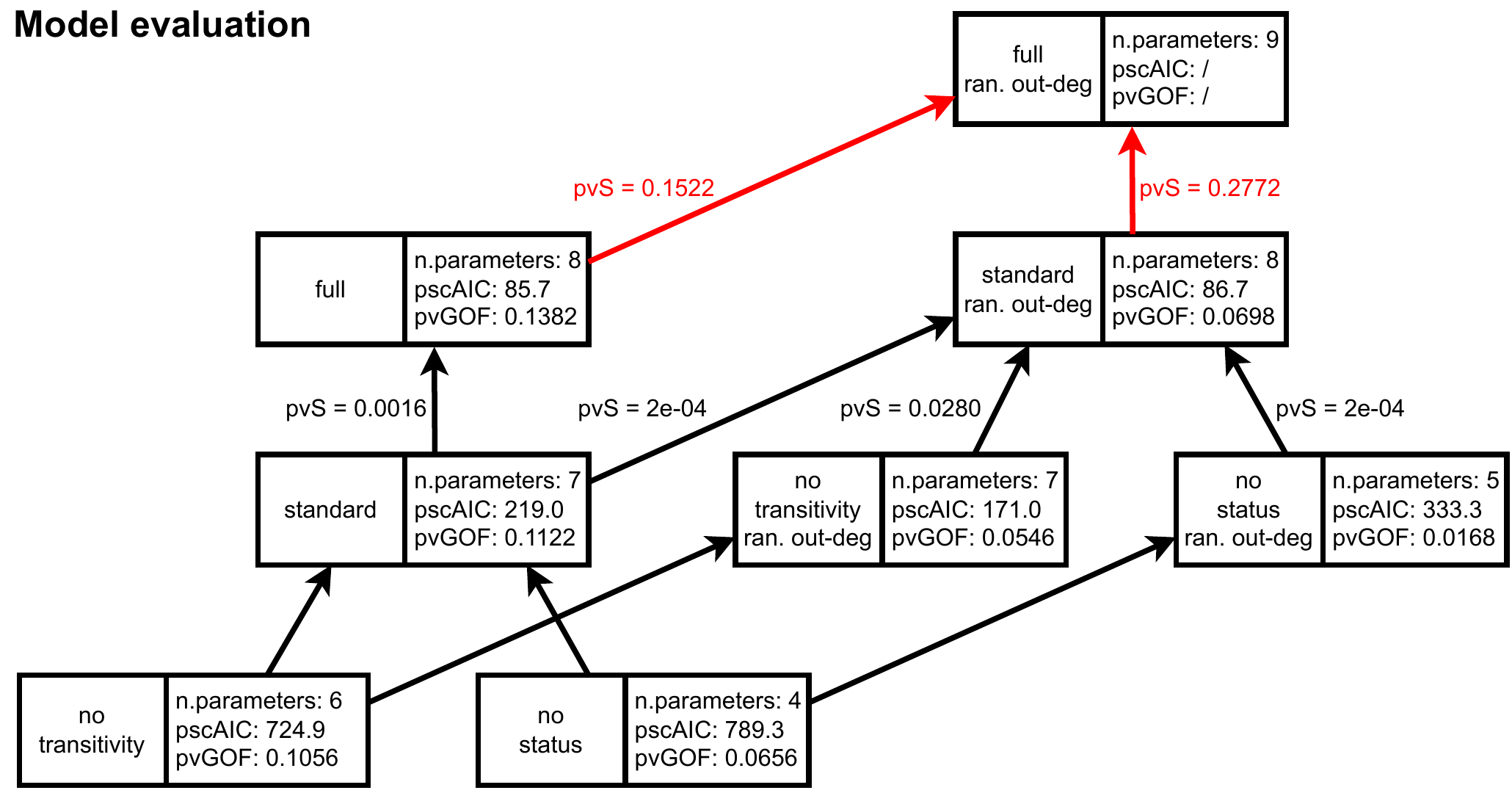}
    \caption{Arrow $\text{\emph{Model 1}}\xrightarrow{}\text{\emph{Model 2}}$ if the parameter space of \emph{Model 1} is a subset of the one of \emph{Model 2}. Some of the arrows were formally tested, with null hypothesis $H_0:\text{\emph{Model 1 is true model}}$, and associated empirical \emph{p}-values \emph{pvS}. For each model are plotted the number of parameters, the AIC parameter selection criterion (\emph{pscAIC}) and the \emph{p}-value for the goodness of fit test \emph{pvGOF}. Red arrows denote the score tests in which the null is accepted at 0.05 signifincance level.}
    \label{figure:modelinclusion}
\end{figure}

In this section, we compare different models using the theory developed in Section~\ref{section:ModelEvaluationComputationStandardErrors}.
Figure~\ref{figure:modelinclusion} describes the nesting relations between the eight models, where arrows point from models to models that are their extensions.
If two models are connected by a path of arrows, the null hypothesis that the smaller model is the generative model can be tested.
These are computed using 5,000 simulations from the estimated model under the null hypothesis.
For each of the estimated model, a parameter selection criterion, and a test for goodness of fit is computed.

\subsubsection*{Model comparisons with score tests}
Six score tests are discussed in detail, the empirical \emph{p}-values $\hat{\alpha}$ of these tests are shown in Figure~\ref{figure:modelinclusion}.

\begin{itemize}
    \item \emph{1. Relevance of random out-degree in the standard model.} We test the null hypothesis that there is no random out-degree effect, $H_0:\sigma^2=0$, versus the alternative $H_1:\sigma^2>0$.
    The empirical \emph{p}-value computed with equation (\ref{eq:empiricalp}) is equal to $\hat{\alpha}=2\cdot 10^{-4}$, whereas the asymptotic one computed with (\ref{equation:chisqtest2}) is lower than $1/5000 = 2\cdot 10^{-4}$.
Note that in the right-hand plot of Figure~\ref{figure:algorithm}, the observed value of the out-degree variance aligns moderately well with the distribution of the simulated variances of the out-degrees obtained from Phase~3 (the proportion of higher simulated values is 0.09).
However, this correspondence does not take into account that parameters are estimated.
This shows the higher power associated with the orthogonalization.

\item \emph{2. Irrelevance of random out-degree in the full model.} Similarly, we test the null hypothesis of no random out-degree in the full model.
The \emph{p}-values computed with the empirical and asymptotic distribution are $\hat{\alpha}=0.27$ and $\tilde{\alpha}=0.44$, respectively, so there is no evidence against the null hypothesis in this model.
This is expected because when activity and overdispersion in the out-degree are both included, the model becomes almost no-identifiable.

\item \emph{3. Relevance of out-degree activity effect for the model without random out-degrees.} We test the null hypothesis of the out-degree activity effect, $H_0:\beta_{\mbox{\scriptsize out-d. activity}}=0$, against the two-sided alternative  $\beta_{\mbox{\scriptsize out-d. activity}}\ne 0$.
The empirical \emph{p}-value for this test is $\hat{\alpha}=1.16\cdot 10^{-3}$, and $\tilde{\alpha}=4.47\cdot 10^{-4}$ for the test that relies on the normal approximation.
Therefore, if the random out-degree is not included in the model, the out-degree activity effect should be included to explain the observed network dynamics.

\item \emph{4. Irrelevance of out-degree activity effect for the model with random out-degrees.} Similarly, we test the null hypothesis $H_0:\beta_{\mbox{\scriptsize out-d. activity}}=0$ in the model with random out-degrees.
The \emph{p}-values obtained without and with the assumption of normality are $\hat{\alpha}=0.15$ and $\tilde{\alpha}=0.22$, respectively. Therefore, if the random out-degree is  included in the model, the out-degree activity effect does not need to be included to explain the observed data.
This is coherent with test \emph{2}.

\item \emph{5. Relevance at significance level 0.05 of transitivity for the model with random out-degrees.} As discussed in the previous section, the inclusion of the random out-degree reduces the estimated parameter for the transitive triplets. Therefore we test the null hypothesis $H_0:\beta_{\mbox{\scriptsize tr. triplets}} = 0$. Empirical and asymptotic \emph{p}-values are $2.80\cdot 10^{-2}$ and $2.14\cdot 10^{-2}$, respectively.
Therefore, at significance level 0.05 the null hypothesis is rejected, but it might be retained with lower significance levels.

\item \emph{6. Relevance of status for the model with random out-degrees.} We would like to see whether the heterogeneity in the out-degree is able to account for the effect of status, testing the hypothesis $H_0: \beta_{\mbox{\scriptsize status alter}} = \beta_{\mbox{\scriptsize status ego}} = \beta_{\mbox{\scriptsize status similarity}}=0$.
This is the only composite hypothesis tested.
The empirical \emph{p}-value is $\hat{\alpha}=2\cdot 10^{-4}$, whereas the asymptotic one is lower than $2\cdot 10^{-4}$.
So, clearly the random out-degree is not able to capture fully the effect of status.
\end{itemize}

As a result, the model containing both out-degree activity and random out-degree is over-parametrized, whereas the model that does not contain these two effects is under-parametrized.
The random out-degree is not able to account for the effect that the status of the individuals has on the social dynamics.
The null hypothesis of statistical equivalence between the models without and with transitive triplets is rejected at significance level 0.05, although it would be accepted at significance level 0.01, suggesting that the random out-degree is partially responsible of some of the observed triangles in the network.

\subsubsection*{Model comparison with parameter selection criterion}
For all models but the full with random out-degree (which has not been estimated), the parameter selection criterion defined in equation (\ref{eq:psc}) has been computed.
The set of statistics $\bm{g}^0$ that is used to evaluate the models is the 9 dimensional vector containing all statistics present in at least one of the estimated models.
The ``degrees of freedom" function $N\mapsto \mathrm{df}(N)$ has been set to the identity, that is $\mathrm{df}(N) = N$.
The reason is that social networks are sparse, meaning that the number of ties (and so of sub-graphs counts used as statistics) grows linearly with the size of the network.
Other alternatives, such as $\mathrm{df}(N) = N(N-1)$  have been tried, but the comparison between the seven estimated models is not changed because all network have the same size, and the first component of the parameter selection criterion is the dominant one.
In fact, AIC and BIC information criteria also lead to the same ``best model", so only the former is given, for all estimated models, in Figure~\ref{figure:modelinclusion}.

The full model without random effects is the one that minimizes the AIC parameter selection criterion, that in this case is equal to $85.66$.
However, the AIC of the standard model with random effects is equal to $86.71$, so it is only slightly higher than the previous one.
All other estimated models have a substantially higher AIC, so they are much worse.
Note that the inclusion of the random out-degree decreases significantly the AIC of all models, except for the full model that is not identifiable when the random effect is included.
In particular, the no-transitivity model with random out-degree has a lower AIC than the standard model without the random effect.
This fact supports the claim that the overdispersion of the out-degree partially explains the triangles in the model.
However, the inclusion of both random out-degree and transitive triplets effects strongly reduce the parameter selection criterion, suggesting that both effects should be included in the model, when out-degree activity is not used.

\subsubsection*{Goodness of fit with out-degree distribution.}
The \emph{out-degree distribution} 
\begin{equation*}
\bm{a}(\bm{x}) \propto (|\text{nodes with out-degree }k|)_{k=0,...,20}\in [0,1)^{21}, 
\end{equation*}
that is the vector containing the proportion of nodes with out-degree equal to $0$, $1$, ..., $20$, has been used as auxiliary statistics. 
This vector is standardized so that the sum of its components is equal to 1.
The \emph{p}-value for the goodness of fit test computed with equation (\ref{eq:pvalueGOF}) requires the inversion of the matrix $\hat{\bm{\Omega}}$, that is the approximated variance of the auxiliary statistic.
But $\hat{\bm{\Omega}}$ is not invertible because there are some values for the out-degree in which the proportion of nodes with this out-degrees is zero in all simulations.
Therefore $\hat{\bm{\Omega}}$ is corrected by adding $0.2$ to its diagonal components.
We have checked that varying this correction factor and the maximum out-degree $20$ that is considered to compute the auxiliary statistics, almost does not change the \emph{p}-values.

The \emph{p}-values for the goodness of fit test in all estimated models are denoted as \emph{pvGOF} in Figure~\ref{figure:modelinclusion}.
All models with random effects have lower \emph{p}-values than the correspondent ones without the random effects.
However, these \emph{p}-values should not be used to compare different models, like the parameter selection criterion discussed before, but they can only be used to reject the null in which the model does not fit the data.
At significance level 0.05 only the no-status model with random out degree would be rejected with this method.
But at higher significance level more models would be rejected.

Other auxiliary statistics could have been used instead of the out-degree distribution. 
Important examples are the \emph{in-degree distribution}, which is related with the \emph{attractiveness bias} described in \cite{feld2002detecting}, or the \emph{triad census}, which contains information about the local structure of transitivity.
These and other auxiliary statistics are described and compared extensively in \cite{lospinoso2019goodness}.

\subsection{Out-degree activity and overdispersion in Tailor shop network}

The main purpose of the statistical analysis was the illustration of the theory developed in the previous sections in an empirical dataset.
The emphasis has been on the dichotomy between modelling the observed variability of the out-degree distribution including either the out-degree activity effect or the random out-degree effect.
These two differ in their interpretation as drivers for social behaviour of the actors when forming and removing ties.
The choice of effect also changes the estimates and the standard deviations of the other parameters in the evaluation function.
The model containing both effects included is overparametrized, and the estimation algorithm does not converge because of identifiability issues.
The two models that emerge as the best ones from the estimating and testing procedures are not nested and therefore cannot be formally compared by a score test.
However, the two models can be compared with the AIC parameter selection criterion, but even in this case, there is not a strong preference for one of the two models.

The interpretation of a positive parameter for the out-degree activity effect is that actors with a larger out-degree are more likely to form ties.
The interpretation of the random out-degree, on the other hand, is unobserved heterogeneity between actors in their tendency to create outgoing ties.
The former model provides an explanation for the observed variability together with a homogeneity assumption. 
However, the parameter information criterion is slightly lower for the model with out-degree activity, so the choice between the two models should be made on conceptual grounds.

The random out-degree cannot replace the status of the individuals, as models without interaction effects with this covariate have a much worse fit. 
However, the random out-degree effect can partially account for the observed triangles in the network. 
The reason is that some of the triangles are closed by chance, while others are due to the fact that individuals like to form and be part of triangles in their social relationships.
The score test for statistical equivalence between the no-transitivity and the standard model with random effects has a \emph{p}-value between 0.01 and 0.05, but the parameter selection criterion strongly suggests that transitive triplets should be included in the model.
Therefore, even though the triangles in the model can be partially explained by the random out-degree, not including the transitive triplets causes a worse fit overall.

\section{Overview and discussion}
\label{section:Discussion}
In this paper we have generalized the stochastic actor oriented model to include random effects in the evaluation function. This generalization overcomes the limitation that all actors in the study must have the same parameters, allowing actors to react in their own idiosyncratic way to respond to certain circumstances. The generalization is important because random effects are commonly used in longitudinal studies, as the main method to parametrize the heterogeneity of the actors in the study when no suitable covariates are available.

The consideration of using a SAOM with random effects depends on a number of factors. First, it involves identifying covariates relevant to answering the research question of interest, as well as additional variables to control for confounding. In our generalization, some of these included effects can also be random. This means that the effect for a certain covariate may vary according to the individual experiencing it. Secondly, parameters are estimated via an adapted simulated method of moments procedure. In these simulations, we also compute all the quantities needed for model evaluation. This includes score tests to determine (i) whether a fixed effect is statistically significant, or, in our proposed generalization, (ii) whether individual heterogeneity expressed through a random effect is really needed or not. Thirdly, goodness-of-fit and parameter selection criteria can also be used to assess the overall fit and to compare non-nested models. Finally, the parameters associated with effects relevant to the research question are interpreted. If various models fit the data equally well, they can be selected based on the interpretations of their parameters, or the emphasis can be on interpreting the results of various models with a good fit, as they may highlight different aspects of the underlying social dynamics.

The distribution of the random parameters is parametrized by their variance, which is shared by all actors, and it is estimated with the method of moments, as are the other global parameters. This estimation method is less time-consuming than likelihood-based methods. The generalizations of the evaluation function presented in Section~\ref{section:SAOMRandomOutEffects}, and of the estimation algorithm, described in Section~\ref{sec:estimation}, were relatively straightforward.
In Section~\ref{section:ModelEvaluationComputationStandardErrors} score-type tests for comparing nested models are discussed, as well as how standard errors are computed when there are random effects, and other generalizations including a goodness of fit test and a model selection procedure that can be used to compare non-nested models. In Section~\ref{section:EstimationKapferer} we fitted and compared various models, with and without random out-degree, on Kapferer's (\citeyear{kapferer1972strategy}) tailor shop data. In some models, some important effects were purposely left out, to test whether the random effect can partially overcome the excluded information. We found that when important information is left out, the estimated variance of the out-degree parameter is higher, meaning that heterogeneity of actors is one way to fit the observed network data. Two of the models considered explain the observed data significantly better than the others. The first one includes the out-degree activity effect and has no random effects. The second model includes the random out-degree effect, and not the out-degree activity effect. We cannot formally test which of the two models explains the data better because they are not nested, and the parameter selection criterion does not provide a definite answer. Therefore, the choice between them must depend on substantive theory and the preferred interpretation.

In the analyzed dataset, the inclusion of the random effect increases the variability of the out-degree distribution, resulting in a higher occurrence of coincidental closure of triangles.
Consequently, a lower parameter for the transitive triplets effect is needed to account for the disparity between observed and simulated transitive triplets.
Nevertheless, in our case, this parameter is statistically significant.
Hence, the overdispersion in the modeled out-degree distribution, governed by the random out-degree, cannot act as a substitute for the transitive triple effect.
We expect that random dyadic effects may capture a portion of degree closure, even in larger social networks segmented into multiple communities. 
However, their capacity is limited to regulating the chance closure of triangles, but it is known that relationships between individuals such as friendship, tend to form triangles.
Therefore transitivity effects are still important, especially in large networks.
Despite this, random dyadic effects can still be useful in discerning which transitivity effects should be incorporated into the model, particularly in directed networks, where numerous alternatives exist for modeling transitivity.

Although the discussion of the Kapferer study constituted an elaborate example of a model with a random out-degree effect, other random effects varying between actors could also be postulated. This goes beyond the scope of the current manuscript, but would be interesting to explore. As our method is very general in principle, it would be useful to further elaborate on models with multiple correlated random effects, as mentioned in Section~\ref{sec:generalizationsOfModelEvaluation}. Currently, the implementation discussed in Appendix~\ref{sec:implementation dummy variables} can be used to simulate and estimate models with random effects. By extending the code of \emph{RSiena}, the computational complexity of simulating a network from a model with or without random effects will be the same, up to a factor that is constant with respect to the size of the network. The number of iterations needed to approximate all quantities required for model evaluation may be higher for models with random effects, because the sampling of the random parameters introduces a new source of variability. However, the difference in the number of iterations should not depend on the size of the network.

The stochastic actor oriented model with random effects provides the flexibility to choose which parameters should be global, thus common to all actors, and which ones are random, so as to allow heterogeneity between actors.
We have described an efficient score test for the presence of a random effect and a goodness of fit procedure, extending the approaches for fixed effects.
A new parameter selection criterion has been also developed.
Therefore the researcher can evaluate different models, and choose the ones that are theoretically and empirically most appealing.

\section*{Acknowledgement}
Giacomo Ceoldo and Ernst C. Wit acknowledge funding from the \emph{Swiss National Science Foundation} (SNSF 188534).

\bibliography{biblio}

\begin{thebibliography}{50}
\providecommand{\natexlab}[1]{#1}
\providecommand{\url}[1]{\texttt{#1}}
\expandafter\ifx\csname urlstyle\endcsname\relax
  \providecommand{\doi}[1]{doi: #1}\else
  \providecommand{\doi}{doi: \begingroup \urlstyle{rm}\Url}\fi

\bibitem[Amati et~al.(2015)Amati, Sch\"{o}nenberger, and Snijders]{ASS2015}
Viviana Amati, Felix Sch\"{o}nenberger, and Tom A~B Snijders.
\newblock Estimation of stochastic actor-oriented models for the evolution of
  networks by generalized method of moments.
\newblock \emph{Journal de la Soci\'{e}t\'{e} Fran\c{c}aise de Statistique},
  156:\penalty0 140--165, 2015.

\bibitem[Amati et~al.(2019)Amati, Sch\"{o}nenberger, and Snijders]{ASS2019}
Viviana Amati, Felix Sch\"{o}nenberger, and Tom A~B Snijders.
\newblock Contemporaneous statistics for estimation in stochastic
  actor-oriented co-evolution models.
\newblock \emph{Psychometrika}, 84:\penalty0 1068--1096, 2019.

\bibitem[Andrews(1999)]{andrews1999consistent}
Donald~WK Andrews.
\newblock Consistent moment selection procedures for generalized method of
  moments estimation.
\newblock \emph{Econometrica}, 67\penalty0 (3):\penalty0 543--563, 1999.

\bibitem[Basawa(1985)]{basawa1985neyman}
IV~Basawa.
\newblock Neyman-le cam tests based on estimating functions.
\newblock In \emph{Proceedings of the Berkeley conference in honor of Jerzy
  Neyman and Jack Kiefer}, volume~2, pages 811--825. Wadsworth Belmont, Calif,
  USA, 1985.

\bibitem[Basawa(1991)]{basawa1991generalized}
IV~Basawa.
\newblock Generalized score tests for composite hypotheses.
\newblock \emph{Estimating functions}, pages 121--131, 1991.

\bibitem[Berkhof and Snijders(2001)]{berkhof2001variance}
Johannes Berkhof and Tom A~B Snijders.
\newblock Variance component testing in multilevel models.
\newblock \emph{Journal of Educational and Behavioral Statistics}, 26\penalty0
  (2):\penalty0 133--152, 2001.

\bibitem[Block et~al.(2018)Block, Koskinen, Hollway, Steglich, and
  Stadtfeld]{block2018change}
Per Block, Johan Koskinen, James Hollway, Christian Steglich, and Christoph
  Stadtfeld.
\newblock Change we can believe in: Comparing longitudinal network models on
  consistency, interpretability and predictive power.
\newblock \emph{Social Networks}, 52:\penalty0 180--191, 2018.

\bibitem[Block et~al.(2019)Block, Stadtfeld, and Snijders]{block2019forms}
Per Block, Christoph Stadtfeld, and Tom A~B Snijders.
\newblock Forms of dependence: Comparing saoms and ergms from basic principles.
\newblock \emph{Sociological Methods \& Research}, 48\penalty0 (1):\penalty0
  202--239, 2019.

\bibitem[Butts(2008)]{butts20084}
Carter~T Butts.
\newblock 4. a relational event framework for social action.
\newblock \emph{Sociological Methodology}, 38\penalty0 (1):\penalty0 155--200,
  2008.

\bibitem[Caimo et~al.(2014)Caimo, Friel, et~al.]{caimo2014bergm}
Alberto Caimo, Nial Friel, et~al.
\newblock Bergm: Bayesian exponential random graphs in {R}.
\newblock \emph{Journal of Statistical Software}, 61\penalty0 (i02), 2014.

\bibitem[Carley(1990)]{carley1990group}
Kathleen~M Carley.
\newblock Group stability: A socio-cognitive approach.
\newblock \emph{Advances in Group Processes}, 7\penalty0 (1):\penalty0 44,
  1990.

\bibitem[DuBois et~al.(2013)DuBois, Butts, McFarland, and
  Smyth]{dubois2013hierarchical}
Christopher DuBois, Carter~T Butts, Daniel McFarland, and Padhraic Smyth.
\newblock Hierarchical models for relational event sequences.
\newblock \emph{Journal of Mathematical Psychology}, 57\penalty0 (6):\penalty0
  297--309, 2013.

\bibitem[Feld and Carter(2002)]{feld2002detecting}
Scott~L Feld and William~C Carter.
\newblock Detecting measurement bias in respondent reports of personal
  networks.
\newblock \emph{Social networks}, 24\penalty0 (4):\penalty0 365--383, 2002.

\bibitem[Handcock(2003)]{handcock2003statistical}
Mark~S Handcock.
\newblock Statistical models for social networks: degeneracy and inference.
\newblock \emph{Dynamic Social Network Modeling and Analysis. National
  Academies Press, Washington, DC}, pages 229--240, 2003.

\bibitem[Hanneke et~al.(2010)Hanneke, Fu, and Xing]{hanneke2010discrete}
Steve Hanneke, Wenjie Fu, and Eric~P Xing.
\newblock Discrete temporal models of social networks.
\newblock \emph{Electronic Journal of Statistics}, 4:\penalty0 585--605, 2010.

\bibitem[Higham(1989)]{higham1988matrix}
Nicholas~J Higham.
\newblock Matrix nearness problems and applications.
\newblock In M~J~C Gover and S~Barnett, editors, \emph{Applications of Matrix
  Theory}, pages 1--27. Oxford University Press, Oxford, 1989.

\bibitem[Holland and Leinhardt(1977{\natexlab{a}})]{holland1977dynamic}
Paul~W Holland and Samuel Leinhardt.
\newblock A dynamic model for social networks.
\newblock \emph{Journal of Mathematical Sociology}, 5\penalty0 (1):\penalty0
  5--20, 1977{\natexlab{a}}.

\bibitem[Holland and Leinhardt(1977{\natexlab{b}})]{holland1977method}
Paul~W Holland and Samuel Leinhardt.
\newblock A method for detecting structure in sociometric data.
\newblock \emph{Social Networks}, pages 411--432, 1977{\natexlab{b}}.

\bibitem[Kapferer(1972)]{kapferer1972strategy}
Bruce Kapferer.
\newblock \emph{Strategy and transaction in an African factory: African workers
  and Indian management in a Zambian town}.
\newblock Manchester University Press, 1972.

\bibitem[Koskinen and Snijders(2007)]{KoskinenSnijders07}
Johan~H. Koskinen and Tom A.~B. Snijders.
\newblock Bayesian inference for dynamic social network data.
\newblock \emph{Journal of Statistical Planning and Inference}, 13:\penalty0
  3930--3938, 2007.

\bibitem[Krivitsky and Handcock(2014)]{krivitsky2014separable}
Pavel~N Krivitsky and Mark~S Handcock.
\newblock A separable model for dynamic networks.
\newblock \emph{Journal of the Royal Statistical Society: Series B (Statistical
  Methodology)}, 76\penalty0 (1):\penalty0 29--46, 2014.

\bibitem[Lospinoso and Snijders(2019)]{lospinoso2019goodness}
Josh Lospinoso and Tom A~B Snijders.
\newblock Goodness of fit for stochastic actor-oriented models.
\newblock \emph{Methodological Innovations}, 12\penalty0 (3):\penalty0
  2059799119884282, 2019.

\bibitem[Mood(2010)]{mood2010logistic}
Carina Mood.
\newblock Logistic regression: Why we cannot do what we think we can do, and
  what we can do about it.
\newblock \emph{European Sociological Review}, 26:\penalty0 67--82, 2010.

\bibitem[Neyman(1959)]{neyman1959optimal}
Jerzy Neyman.
\newblock Optimal asymptotic tests of composite hypotheses.
\newblock In Ulf Grenander, editor, \emph{Probability and Statistics}, pages
  213--234. Wiley, 1959.

\bibitem[Nowicki and Snijders(2001)]{nowicki2001estimation}
Krzysztof Nowicki and Tom A~B Snijders.
\newblock Estimation and prediction for stochastic blockstructures.
\newblock \emph{Journal of the American Statistical Association}, 96\penalty0
  (455):\penalty0 1077--1087, 2001.

\bibitem[Perry and Wolfe(2013)]{perry2013point}
Patrick~O Perry and Patrick~J Wolfe.
\newblock Point process modelling for directed interaction networks.
\newblock \emph{Journal of the Royal Statistical Society: SERIES B: Statistical
  Methodology}, pages 821--849, 2013.

\bibitem[Polyak and Juditsky(1992)]{polyak1992acceleration}
Boris~T Polyak and Anatoli~B Juditsky.
\newblock Acceleration of stochastic approximation by averaging.
\newblock \emph{SIAM journal on control and optimization}, 30\penalty0
  (4):\penalty0 838--855, 1992.

\bibitem[Ripley et~al.(2023)Ripley, Snijders, Boda, V{\"o}r{\"o}s, and
  Preciado]{manualRsiena}
Ruth~M Ripley, Tom A~B Snijders, Zs{\'o}fia Boda, Andr{\'a}s V{\"o}r{\"o}s, and
  Paulina Preciado.
\newblock Manual for rsiena.
\newblock \emph{University of Oxford, Department of Statistics, Nuffield
  College}, 2023.

\bibitem[Robbins and Monro(1951)]{robbins1951stochastic}
Herbert Robbins and Sutton Monro.
\newblock A stochastic approximation method.
\newblock \emph{The Annals of Mathematical Statistics}, 22:\penalty0 400--407,
  1951.

\bibitem[Robins et~al.(2004)Robins, Pattison, and Woolcock]{robins2004missing}
Garry Robins, Philippa Pattison, and Jodie Woolcock.
\newblock Missing data in networks: exponential random graph (p*) models for
  networks with non-respondents.
\newblock \emph{Social Networks}, 26\penalty0 (3):\penalty0 257--283, 2004.

\bibitem[Rubinstein(1989)]{rubinstein1989sensitivity}
Reuven~Y Rubinstein.
\newblock Sensitivity analysis and performance extrapolation for computer
  simulation models.
\newblock \emph{Operations Research}, 37\penalty0 (1):\penalty0 72--81, 1989.

\bibitem[Rudin(1964)]{rudin1964principles}
Walter Rudin.
\newblock \emph{Principles of Mathematical Analysis}, volume~3.
\newblock McGraw-Hill New York, 1964.

\bibitem[Schweinberger(2007)]{schweinberger2007random}
Michael Schweinberger.
\newblock \emph{Statistical methods for studying the evolution of networks and
  behavior}.
\newblock PhD thesis, Rijksuniversiteit Groningen, 2007.

\bibitem[Schweinberger(2012)]{schweinberger2012statistical}
Michael Schweinberger.
\newblock Statistical modelling of network panel data: Goodness of fit.
\newblock \emph{British Journal of Mathematical and Statistical Psychology},
  65\penalty0 (2):\penalty0 263--281, 2012.

\bibitem[Schweinberger(2020)]{schweinberger2020statistical}
Michael Schweinberger.
\newblock Statistical inference for continuous-time markov processes with block
  structure based on discrete-time network data.
\newblock \emph{Statistica Neerlandica}, 74\penalty0 (3):\penalty0 342--362,
  2020.

\bibitem[Schweinberger and Snijders(2007)]{schweinberger2007markov}
Michael Schweinberger and Tom A~B Snijders.
\newblock Markov models for digraph panel data: {Monte Carlo}-based derivative
  estimation.
\newblock \emph{Computational Statistics \& Data Analysis}, 51:\penalty0
  4465--4483, 2007.

\bibitem[Snijders(1996)]{snijders1996stochastic}
Tom A~B Snijders.
\newblock Stochastic actor-oriented models for network change.
\newblock \emph{Journal of Mathematical Sociology}, 21\penalty0 (1-2):\penalty0
  149--172, 1996.

\bibitem[Snijders(2001)]{snijders2001statistical}
Tom A~B Snijders.
\newblock The statistical evaluation of social network dynamics.
\newblock \emph{Sociological Methodology}, 31:\penalty0 361--395, 2001.

\bibitem[Snijders(2017)]{snijders2017stochastic}
Tom A~B Snijders.
\newblock Stochastic actor-oriented models for network dynamics.
\newblock \emph{Annual Review of Statistics and Its Application}, 4:\penalty0
  343--363, 2017.

\bibitem[Snijders(2023)]{SieAlg}
Tom A~B Snijders.
\newblock Siena algorithms.
\newblock Technical report, University of Groningen, University of Oxford,
  \newline{\small\url{http://www.stats.ox.ac.uk/~snijders/siena/Siena_algorithms.pdf}},
  2023.

\bibitem[Snijders et~al.(2007)Snijders, Steglich, and
  Schweinberger]{snijders2007no}
Tom A.~B. Snijders, Christian E.~G. Steglich, and Michael Schweinberger.
\newblock Modeling the co-evolution of networks and behavior.
\newblock In Kees van Montfort, Han Oud, and Albert Satorra, editors,
  \emph{Longitudinal models in the behavioral and related sciences}, pages
  41--71. Mahwah, NJ: Lawrence Erlbaum, 2007.

\bibitem[Snijders et~al.(2010)Snijders, Koskinen, and
  Schweinberger]{snijders2010maximum}
Tom A~B Snijders, Johan Koskinen, and Michael Schweinberger.
\newblock Maximum likelihood estimation for social network dynamics.
\newblock \emph{The Annals of Applied Statistics}, 4\penalty0 (2):\penalty0
  567, 2010.

\bibitem[Stadtfeld and Geyer-Schulz(2011)]{stadtfeld2011analyzing}
Christoph Stadtfeld and Andreas Geyer-Schulz.
\newblock Analyzing event stream dynamics in two-mode networks: An exploratory
  analysis of private communication in a question and answer community.
\newblock \emph{Social Networks}, 33\penalty0 (4):\penalty0 258--272, 2011.

\bibitem[Stadtfeld et~al.(2017)Stadtfeld, Hollway, and
  Block]{stadtfeld2017dynamic}
Christoph Stadtfeld, James Hollway, and Per Block.
\newblock Dynamic network actor models: Investigating coordination ties through
  time.
\newblock \emph{Sociological Methodology}, 47\penalty0 (1):\penalty0 1--40,
  2017.

\bibitem[Thiemichen et~al.(2016)Thiemichen, Friel, Caimo, and
  Kauermann]{thiemichen2016bayesian}
Stephanie Thiemichen, Nial Friel, Alberto Caimo, and G{\"o}ran Kauermann.
\newblock Bayesian exponential random graph models with nodal random effects.
\newblock \emph{Social Networks}, 46:\penalty0 11--28, 2016.

\bibitem[Uzaheta et~al.(2023)Uzaheta, Amati, and Stadtfeld]{uzaheta2023random}
Alvaro Uzaheta, Viviana Amati, and Christoph Stadtfeld.
\newblock Random effects in dynamic network actor models.
\newblock \emph{Network Science}, 11\penalty0 (2):\penalty0 249--266, 2023.

\bibitem[Van~Duijn et~al.(2004)Van~Duijn, Snijders, and Zijlstra]{van2004p2}
Marijtje~AJ Van~Duijn, Tom~AB Snijders, and Bonne~JH Zijlstra.
\newblock p2: a random effects model with covariates for directed graphs.
\newblock \emph{Statistica Neerlandica}, 58\penalty0 (2):\penalty0 234--254,
  2004.

\bibitem[Wasserman(2004)]{wasserman2004all}
Larry Wasserman.
\newblock \emph{All of statistics: a concise course in statistical inference}.
\newblock Springer, New York, 2004.

\bibitem[Wasserman(1980)]{wasserman1980analyzing}
Stanley Wasserman.
\newblock Analyzing social networks as stochastic processes.
\newblock \emph{Journal of the American Statistical Association}, 75\penalty0
  (370):\penalty0 280--294, 1980.

\bibitem[Wasserman and Faust(1994)]{wasserman1994social}
Stanley Wasserman and Katherine Faust.
\newblock \emph{Social network analysis: Methods and applications}.
\newblock Cambridge University Press, New York and Cambridge, 1994.

\end{thebibliography}
\newpage

\appendix

\section{Identifiability and convergence of optimization}
\label{sec:identifiability}
\begin{figure}
    \centering
\includegraphics[scale=.59]{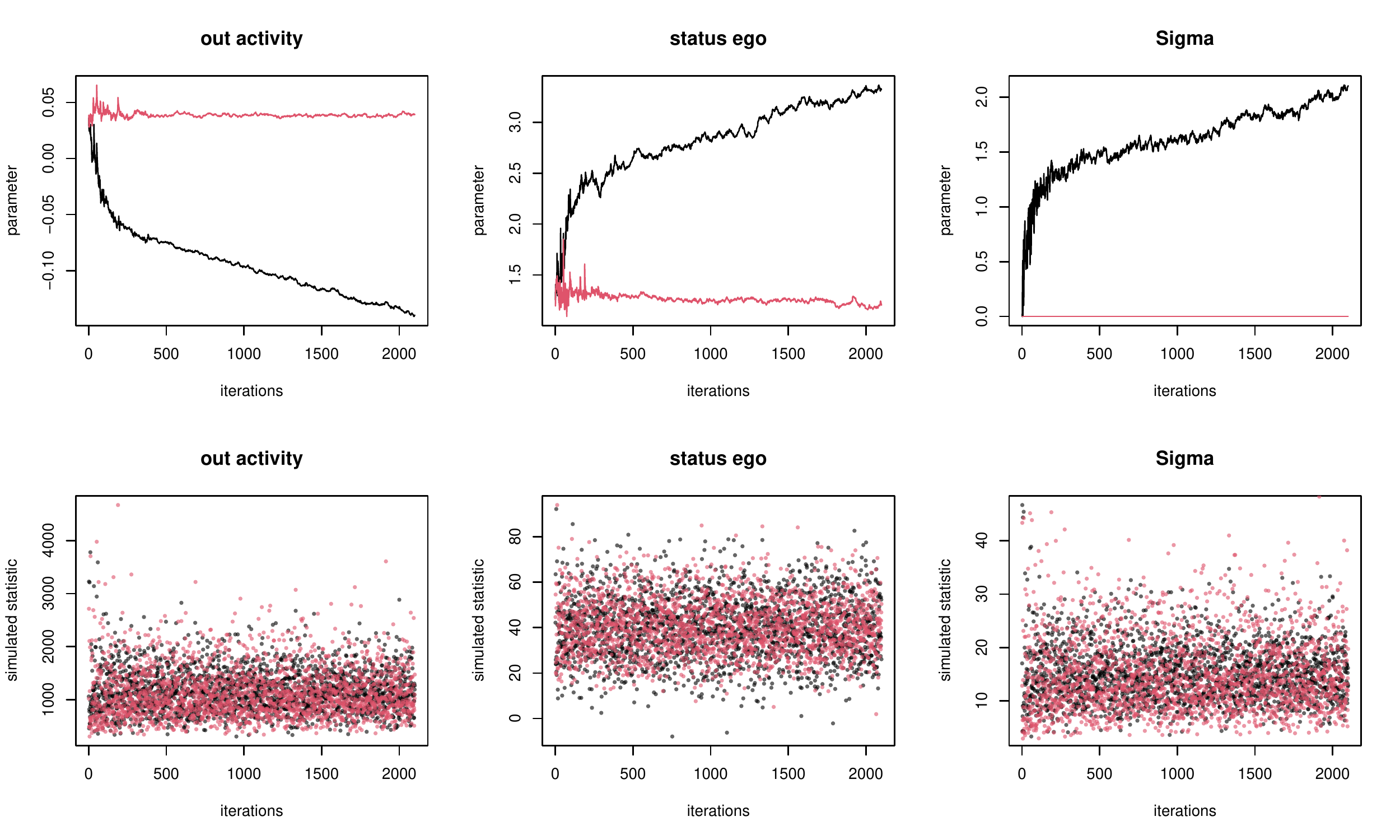}
    \caption{Comparison of chains of parameters (top) and simulated statistics (bottom) for some effects in the full model with (black) and without (red) random out degree.}
    \label{fig:ph2fulltrans}
\end{figure}

In Figure~\ref{fig:ph2fulltrans} it is shown the difference between an estimation in which the moment equation has one solutions, with an optimization in which the model, although theoretically identifiable, cannot be estimated.
In the top row the chains of parameters for some of the effects in the model are shown, in the bottom row the statistics that are simulated from the parameters above are plotted.
In red it is shown the full model without random out-degree described in Section~\ref{section:EstimationKapferer}, and in black the containing the same fixed effects, but with random out-degree also included.
When both out-degree activity and random out-degree are included in the model, the estimation algorithm does not converge, because there are multiple sets of parameters that can simulate statistics equal on average to the observed ones.
In this case, even if the model is theoretically identifiable, as there are not statistics that can be written as linear combination of others, the model is not practically identifiable, because there is not enough information in the data to estimate a model containing out-degree activity, random out-degree and the other effects considered.
Whereas if either one of the out-degree activity or the random out-degree is included, the model can be estimated.

\section{Implementation with dummy variables}
\label{sec:implementation dummy variables}
The simulating function that is currently used in the SAOM can also be used, although in an inefficient way, to simulate models in which the actors can also have individual parameters.
Consider the set of \emph{individual dummy covariates}
\begin{equation}
    \bm{\delta}^{(i)}=(\delta_n^{(i)})_{1\le n\le N} = I(n = i),
\end{equation}
for all actors $i\in\{1,...,N\}$.
The dummy covariates of actor $i$ are ${\delta}^{(i)}_1,...,{\delta}^{(i)}_N$, which are all equal to $0$, except for $\delta_i^{(i)}=1$.
The random out-degree for actor $i$, which is $x_{i+}$, can be then obtained with the \emph{covariate-ego effect} $x_{i+}\delta_i^{(i)}$, which is the interaction between the out-degree $x_{i+}$ and the dummy covariate $\delta_i^{(i)}$.
The latter formula has the advantage that is not only defined for the focal actor $i$, but also for all the other actors, taking value $x_{i+}\delta_n^{(i)} = 0$ when $n\ne i$.
This allows us to ``treat" random parameters $\bm{b}$ as the fixed parameters $\bm{\beta}$ in the simulation, 
so that each actor contains the individual parameter of everyone, but only its own will be relevant in evaluating tie flips, because all other random parameters will be multiplied by an effect equal to 0.
In particular, the evaluation function for actor $i$ is
\begin{equation}
f_{i}(\bm{x})= \beta_1 x_{i+} + \sum_{n=1}^N b_n(x_{i+}\delta_n^{(i)}) +
\sum_{k=2}^p \beta_k s_{ik}(\bm{x}) = (\beta_1 + b_i)x_{i+} + \sum_{k=2}^p \beta_k s_{ik}(\bm{x}),
\end{equation}
which is equivalent to  (\ref{equation:RandomOutDegUtilityFunction}), as $s_{i1}=x_{i+}$.

At iteration $\texttt{t}$, the current variance parameter $\sigma^2_{\texttt{t}}$ is used to simulate $\bm{b}_{\texttt{t}}$, that is then used together with $\bm{\beta}_{\texttt{t}}$ and a set of dummy covariates to simulate the model with the current implementation of \emph{RSiena}.
The function that simulates the process returns the simulated value of the statistic for $\bm{\beta}$, that is the sum of statistics across actors.
With dummy covariates, also the simulated ``individual" statistics are returned, their variance is used to compute the statistic for $\sigma^2$.  
The same procedure can be used when there are more than one random effects.

This implementation is used in the application that is described in Section~\ref{section:EstimationKapferer}.
The code of this implementation is available in \url{https://github.com/gceoldo/SAOM-with-random-effects}.
An important disadvantage of this method is its computational inefficiency, as evaluating every proposed tie flip involves the ``useless" computation of $q(N-1)$ multiplications with an operand equal to $0$, where $q$ is the number of random effects.
This can be avoided, however, in a future \emph{RSiena} implementation.

\section{Mathematical background}

\subsection{Multivariable chain rule}
\label{section:CommDiagramsChainRule}

Consider the functions $f:\mathbb{R}^n\to\mathbb{R}^m$, $g:\mathbb{R}^m\to\mathbb{R}^k$, and their composite $h=(g\circ f):\mathbb{R}^n\to\mathbb{R}^k$.
Let $D_a(h)$ be the \emph{total derivative} of the function $h:\mathbb{R}^n\to\mathbb{R}^k$ evaluated at $a\in\mathbb{R}^n$, i.e.,
\begin{equation}
D_a(h) = \lim_{||\delta||\to 0}
\frac
{|| h(a+\delta) - h(a)||}
{|| \delta ||}.
\end{equation}
The \emph{multivariable chain rule} \citep[Chapter 9]{rudin1964principles} is the formula that relates the total derivative of $h=g\circ f$ with the total derivatives of $f$ and $g$, which is
\begin{equation}
    D_a(h) = D_{f(a)}(g) \circ D_a(f).
\end{equation}
The total derivative is a linear transformation, so the quantities in the last equation can be represented by \emph{Jacobian matrices}, with the composition operation replaced by the matrix product, so that the last equation can be rewritten as
\begin{equation}
    \bm{D}_h(a) = \bm{D}_g(f(a))\cdot \bm{D}_f(a),
    \label{equation:JacobianChainRule}
\end{equation}
where the three matrices are $(k\times n)$, $(k\times m)$ and $(m\times n)$ dimensional, respectively.
Note that the last equation is the generalization of the ``usual" chain rule $h'(a)=g'(f(a))f'(a)$, which is obtained when $n=m=k=1$.

\subsection{Derivative of a Gaussian r.v. with respect to its parameters}
\label{section:derivativesGaussian}
Let $X$ be a Gaussian random variable with mean $\mu$ and variance $\sigma^2$.
Then
\begin{equation}
U=\frac{X-\mu}{\sigma}\sim \mathcal{N}(0,1),
\end{equation}
and $X$ can be written as $
X=\sigma U+\mu \sim \mathcal{N}(\mu,\sigma^2)$.
The standard deviation $\sigma$ is a constant and the distribution of $U$ does not depend on $\mu$, then
\begin{equation}
\frac{\partial X}{\partial \mu} =
\frac{\partial (\sigma U+\mu)}{\partial \mu} = \sigma\frac{\partial U}{\partial\mu} + \frac{\partial\mu}{\partial\mu} \sim \sigma\delta_0+ 1 \sim  \delta_0+1 \sim \delta_1,
\end{equation}
where $\delta_y$ is the degenerate random variable such that $y$ is sampled with probability 1.
Similarly the distribution of $U$ does not depend on $\sigma$, then
\begin{equation}
\frac{\partial X}{\partial \sigma} =
\frac{\partial (\sigma U+\mu)}{\partial \sigma} = \frac{\partial \sigma}{\partial \sigma}U + \frac{\partial \mu}{\partial \sigma} = U \sim \mathcal{N}(0,1),
\end{equation}
and
\begin{equation}
\frac{\partial X}{\partial \sigma^2} =
\frac{\partial ((\sigma^2)^{1/2} U+\mu)}{\partial \sigma^2} = \frac{1}{2}(\sigma^2)^{-1/2}U + 0 = \frac{1}{2\sigma}U \sim \mathcal{N}\Big(0,\frac{1}{4\sigma^2}\Big).
\label{equation:derivativeXwrtsigma2}
\end{equation}

\end{document}